\documentclass[preprint, 3p, twocolumn]{elsarticle}
\usepackage[utf8]{inputenc}
\usepackage{xcolor}
\usepackage{siunitx}
\usepackage{subfig}
\usepackage{caption}
\usepackage[version=4]{mhchem}
\usepackage{textcomp}
\usepackage[normalem]{ulem}

\newcommand{\thgem}{THGEM}
\newcommand{\Vind}{$V_{ind}$}
\newcommand{\Vthgem}{$V_{\thgem}$}
\newcommand{\Vdrift}{$V_{drift}$}

\newcommand{\Ian}{$I_{an}$}
\newcommand{\Itop}{$I_{top}$}
\newcommand{\Ibot}{$I_{bot}$}
\newcommand{\Icath}{$I_{cath}$}
\newcommand{\fullthgem}{FULL}
\newcommand{\rowthgem}{ROW}

\biboptions{sort&compress}

\usepackage{hyperref}

\begin{document}

\begin{frontmatter}

\title{Characterization of a gas detector prototype based on Thick-GEM for the MAGNEX focal plane detector}

\author[1address,2address]{I. Ciraldo}
\author[1address,2address]{G. A. Brischetto\corref{mycorrespondingauthor}}
\cortext[mycorrespondingauthor]{Corresponding author}
\ead{brischetto@lns.infn.it}

\author[2address]{D. Torresi}
\author[2address]{M. Cavallaro}
\author[2address]{C. Agodi}
\author[5address]{A. Boiano}
\author[2address]{S. Calabrese}
\author[1address,2address]{F. Cappuzzello}
\author[2address]{D. Carbone}
\author[6address]{M.Cortesi}
\author[1address,2address,3address]{F. Delaunay}
\author[2address]{M. Fisichella}
\author[2address]{L. Neri}
\author[5address]{A. Pandalone}
\author[5address]{P. Paolucci}
\author[5address]{B. Rossi}
\author[2address]{O. Sgouros}
\author[1address,2address]{V. Soukeras}
\author[1address,2address]{A. Spatafora}
\author[5address]{A. Vanzanella}
\author[4address]{A. Yildirim}
\author[]{\\For the NUMEN collaboration}

\address[1address]{Dipartimento di Fisica e Astronomia ‘‘Ettore Majorana’’, Università di Catania, Italy}
\address[2address]{Laboratori Nazionali del Sud, INFN, Italy}
\address[5address]{Sezione di Napoli, INFN, Italy}
\address[6address]{Facility for Rare Isotope Beams (FRIB), East Lansing, Michigan 48824, USA}
\address[3address]{LPC Caen, Normandie Université, ENSICAEN, UNICAEN, CNRS/IN2P3, Caen, France}
\address[4address]{Department of Physics, Akdeniz University - Antalya, Turkey}

\begin{abstract}
A new gas detector prototype for the upgrade of the focal plane detector of the MAGNEX large-acceptance magnetic spectrometer has been developed and tested in view of the NUMEN project. 
It has been designed to operate at low gas pressure for detecting medium to heavy ions in the energy range between 15 and 60 AMeV. It is a drift chamber based on Multi-layer Thick-GEM (M-\thgem{}) as electron multiplication technology.
Tests with two different M-\thgem{} layouts have been performed using both a radioactive $\alpha$-particle source and accelerated heavy-ion beams. The characterization of the detector in terms of measured currents that flow through the electrodes as a function of different parameters, including applied voltages, gas pressure and rate of incident particle, is described.
The gain and ion backflow properties have been studied.

\end{abstract}


\end{frontmatter}


\section{Introduction}

The NUMEN (NUclear Matrix Elements for Neutrinoless double beta decay) project 
\cite{Cappuzzello2018} proposes an innovative technique to extract information on the nuclear matrix elements 
entering the expression of the lifetime of neutrinoless double-beta decay. 
This is achieved by measuring the cross-sections of specific nuclear reactions such as Double Charge Exchange (DCE) induced by heavy ions \cite{CappuzzelloPPNP}.
The ongoing and planned experiments are performed at INFN - LNS (Italy) 
using the K800 Superconducting Cyclotron and MAGNEX, a large-acceptance magnetic spectrometer \cite{Cappuzzello2015}.

The present tracking system of the MAGNEX Focal Plane Detector (FPD) \cite{Cavallaro2012,Torresi2021} is a large-volume proportional drift chamber working at low pressure, typically in the range 10 $-$ 100 mbar, using wires for the electron multiplication stage.
Such a detector preserves tracking capability up to a rate of a few tens of Hz/cm, with a resolution better than 0.6~mm full width at half maximum (FWHM) in both dispersive (horizontal) and non-dispersive (vertical) coordinates and 5~mrad (FWHM) in horizontal and vertical angles. 
These requirements are fundamental for an accurate particle ray-reconstruction used for the determination of the momentum vector at the target position \cite{Cappuzzello2011}.
In order to measure the extremely small DCE cross sections (of the order of few tens of nb
\cite{Soukeras2021, Cappuzzello2015}) with significant statistics, beam rates up to $10^{13}$~particle per second (pps) at the target are foreseen in the upcoming upgrade of the INFN - LNS infrastructures \cite{Agodi2021}.
With such a high intensity the expected rate of reaction products at the MAGNEX focal plane rises up of more than two orders of magnitude.
A specific upgrade of the MAGNEX FPD is, therefore, mandatory \cite{Cappuzzello2021, Agodi2021, TDR2021,Finocchiaro2020}.

A new gas tracker for the MAGNEX FPD has been designed to satisfy two main requirements: first, it must be able to bear with particle rate at the focal plane of the order of 30 kHz/cm. The second request is that it must preserve the present tracking performance, guaranteeing a sub-millimetric resolution in the measurement of both the horizontal and vertical coordinates, which is essential for an efficient event reconstruction.

A reduced-size prototype of the tracker has been built. The prototype was equipped with a gas-avalanche readout based on the Multi-layer Thick Gas Electron Multiplier (M-\thgem{}) 
\cite{Cortesi2017}.
The M-THGEMs are robust, self-supported, micro-pattern gaseous devices, able to provide sub-millimetric position resolution, good timing properties, and to withstand rates higher than requested, even at low pressure. 
The performance of the detector in terms of gain and ion backflow were computed  by monitoring the currents that flow across all the electrodes of the detector (current mode). The measurements were performed by irradiating the detector effective area with a radioactive $\alpha$-particle source, as well as with a heavy-ion beam.

The paper is organized as follows: the tracker prototype is described in Sect.~\ref{sec:exp_set-up}. 
Sect.~\ref{sec:characterization} reports the characterization of the tracker in terms of measured currents induced on the detector electrodes as a function of the applied voltages, 
gas pressure and rate.  Concluding remarks and outlooks are given in Sect.~\ref{sec:conclusion}.

\section{The gas tracker prototype: operational principle and experimental setup}\label{sec:exp_set-up}

\subsection{The gas tracker prototype}
\label{gas_tracker_prototype}

\begin{figure}[htb]
    \centering
    \includegraphics[scale=0.7]{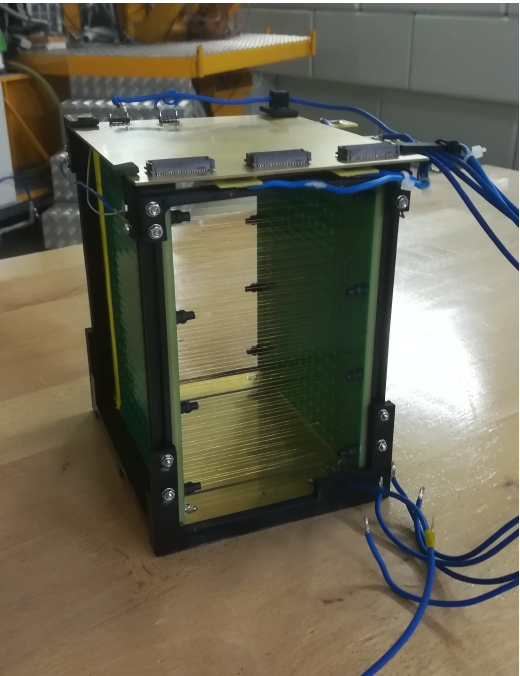}
    \caption{Picture of the gas tracker prototype.}
    \label{fig:photo_prot}
\end{figure}

\begin{figure}[htb]
    \centering
    \includegraphics[width=0.45\textwidth]{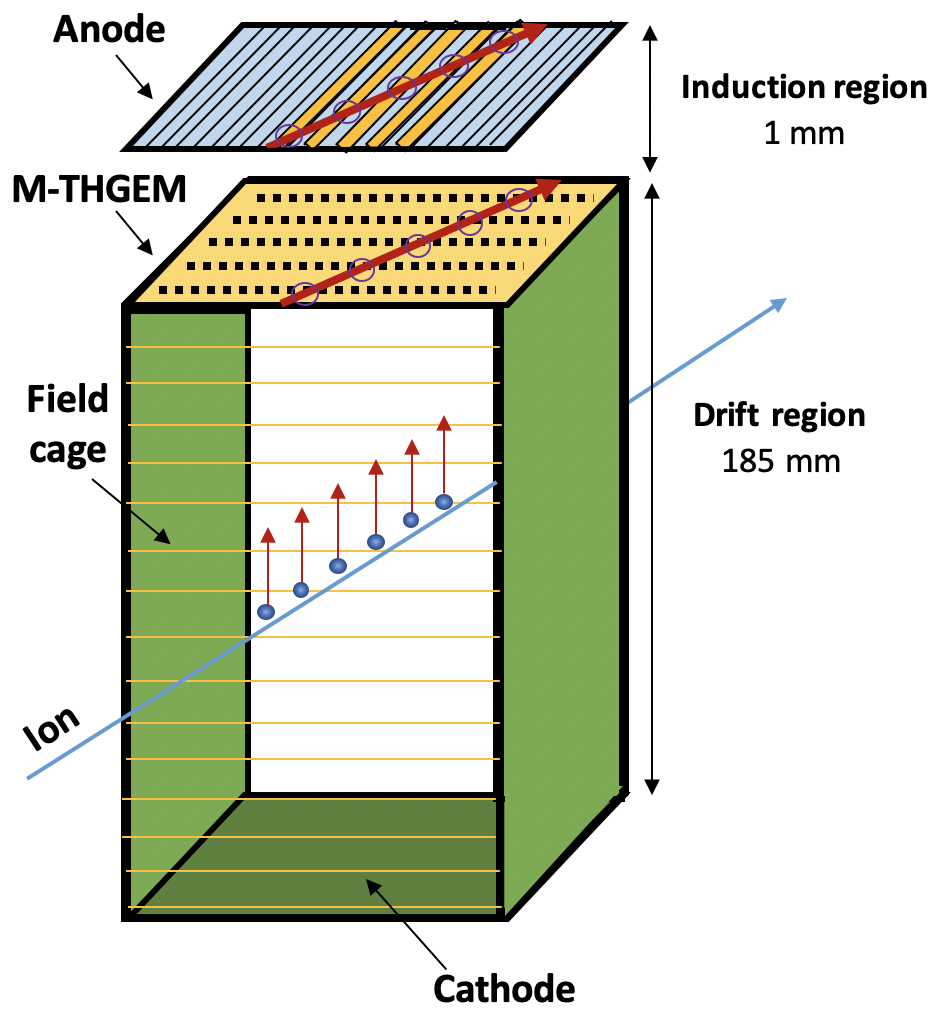}
    \caption{(Color online) Schematic drawing of the gas tracker prototype illustrating the operation principles.}
    \label{fig:operation_principles}
\end{figure}

The gas tracker prototype is a drift
chamber with an active volume of $107 \times 107 \times 185$~$\mbox{mm}^{3}$ (see picture in Fig.~\ref{fig:photo_prot}). It has a smaller size than the foreseen final MAGNEX gas tracker with dimensions of $1200 \times 107 \times 150$~$\mbox{mm}^{3}$. In the present study, the detector was operated in isobutane ($\mbox{iC}_{4}\mbox{H}_{10}$) with 99.95\% purity at pressure typically ranging between a few mbar and several tens of mbar. \\
The volume of the tracker comprises three regions, as sketched in Fig.~\ref{fig:operation_principles}: a drift region, an electron multiplication stage and an induction region.

\paragraph{The drift region}
The drift region is the detector active volume crossed by the incident charged particles, where the primary ionization occurs. It is 185 mm thick and bounded on one side by the cathode, and on the other side by the multiplication stage (M-THGEM). 
A uniform electric field of 50 V/cm is established within the drift region by applying a voltage difference between the cathode and the M-THGEM bottom electrode. A partition grid, made of  50~$\mu m$ thick gold-plated tungsten wires arranged in steps of 5~mm, is used to ensure a good field uniformity across the full volume.

\begin{figure}[h]
	\centering
		\subfloat[] 
		 [\label{fig:row_holes}] {
		 \includegraphics[width=0.42\textwidth]{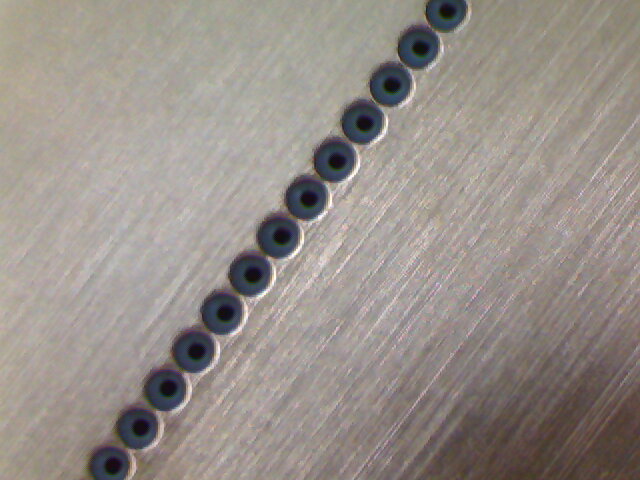}}
		\quad
		\subfloat[]
		 [\label{fig:full_holes}] {
		 \includegraphics[width=0.42\textwidth]{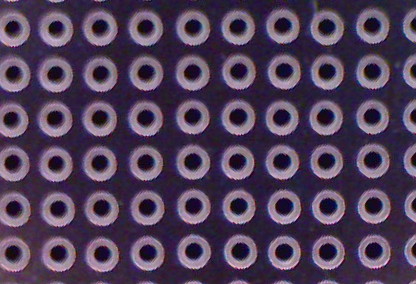}}
	\caption{Magnification of a small area of a ROW M-\thgem{} (a) and a FULL M-\thgem{} (b).}
	\label{fig:thgem_holes}		
\end{figure}

\begin{table*} [htbp]
   \begin{center}
      \caption{Main characteristics of the two types of tested M-\thgem{} foils.}
   \renewcommand{\arraystretch}{1.2}
      \begin{tabular}{|l|c|c|}
        \hline 
		& \textbf{FULL M-THGEM} 	& \textbf{ROW M-THGEM} \\ 
        \hline 
        Substrate material			 & Ceramic SD103K 	& PCB \\ 
        \hline 
        Finish board thickness (mm) & 1.37 	& 1.28 \\ 
        \hline 
        Dimension ($\mbox{mm}^2$) 	 & $107\times 107$ 	& $107 \times 107$ \\ 
        \hline 
        Rim size (mm) 				 & 0.1 				& 0.2  \\ 
        \hline  
        Number of holes					 & 20449 			& 715 \\
        \hline 
        Hole diameter (mm) 		 & 0.30 	& 0.280 \\ 
        \hline
        Hole pitch (mm)				 & 0.75 			& 0.75 \\ 
        \hline
      \end{tabular} 
    \label{tab:thgem_characteristics}
   \end{center}
\end{table*}

\paragraph{The multiplication stage}

The electron multiplication stage is based on M-THGEM
\cite{Shalem2006a,Ayyad2017,Cortesi2020,Cortesi2017,Cortesi2018}. This is a stack of several THGEMs assembled together in a single element. 
THGEMs \cite{Chechik2004,Breskin2009} are micro-pattern gas detectors directly derived from the Gas Electron Multiplier (GEM)~\cite{Sauli2016} but thicker and with a larger hole diameter by a factor between 5 and 50. 
They are produced by multi-layer printed-circuit-board (PCB) technique, mechanically drilling the layers of copper and insulating material laminated together. 
M-THGEMs can provide high gas gain (of the order of $10^6-10^7$ with single-photoelectrons), a rate capability up to $10^8$ Hz/cm$^2$, sub-millimetric position resolution and time resolution of a few~ns.
Moreover, M-THGEMs are a good solution for applications that require to work at low gas pressure~\cite{Shalem2006a,Ayyad2017,Cortesi2020}.
The M-THGEM foils studied in the present work are three-layer THGEMs.

\begin{figure*}[htbp]
	\centering
		\subfloat[]
		 [\label{fig:COMSOL_ROW1}] {
		 \includegraphics[width=0.5\textwidth]{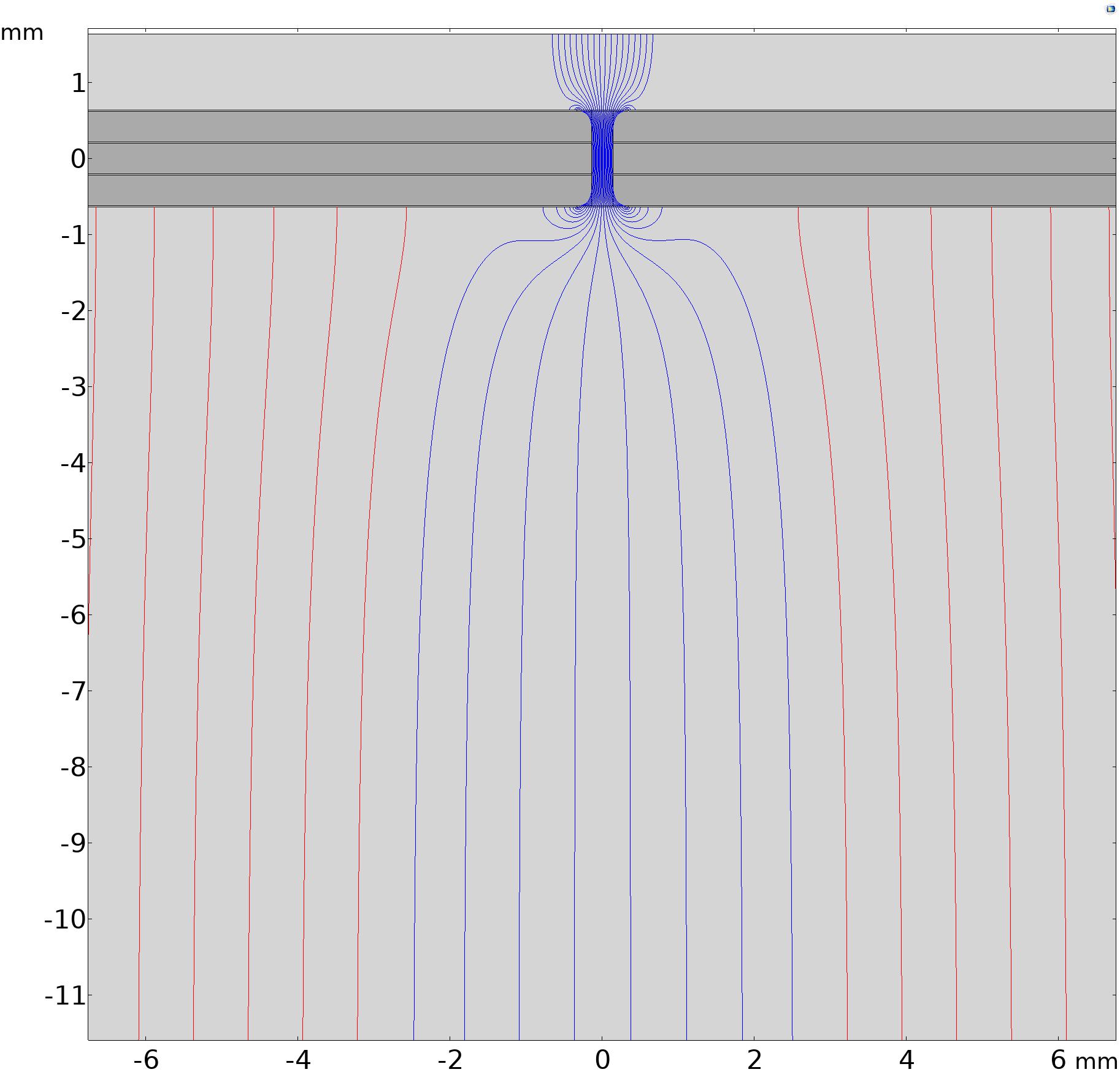}}
		\subfloat[]
		 [\label{fig:COMSOL_ROW2}] {
		 \includegraphics[width=0.5\textwidth]{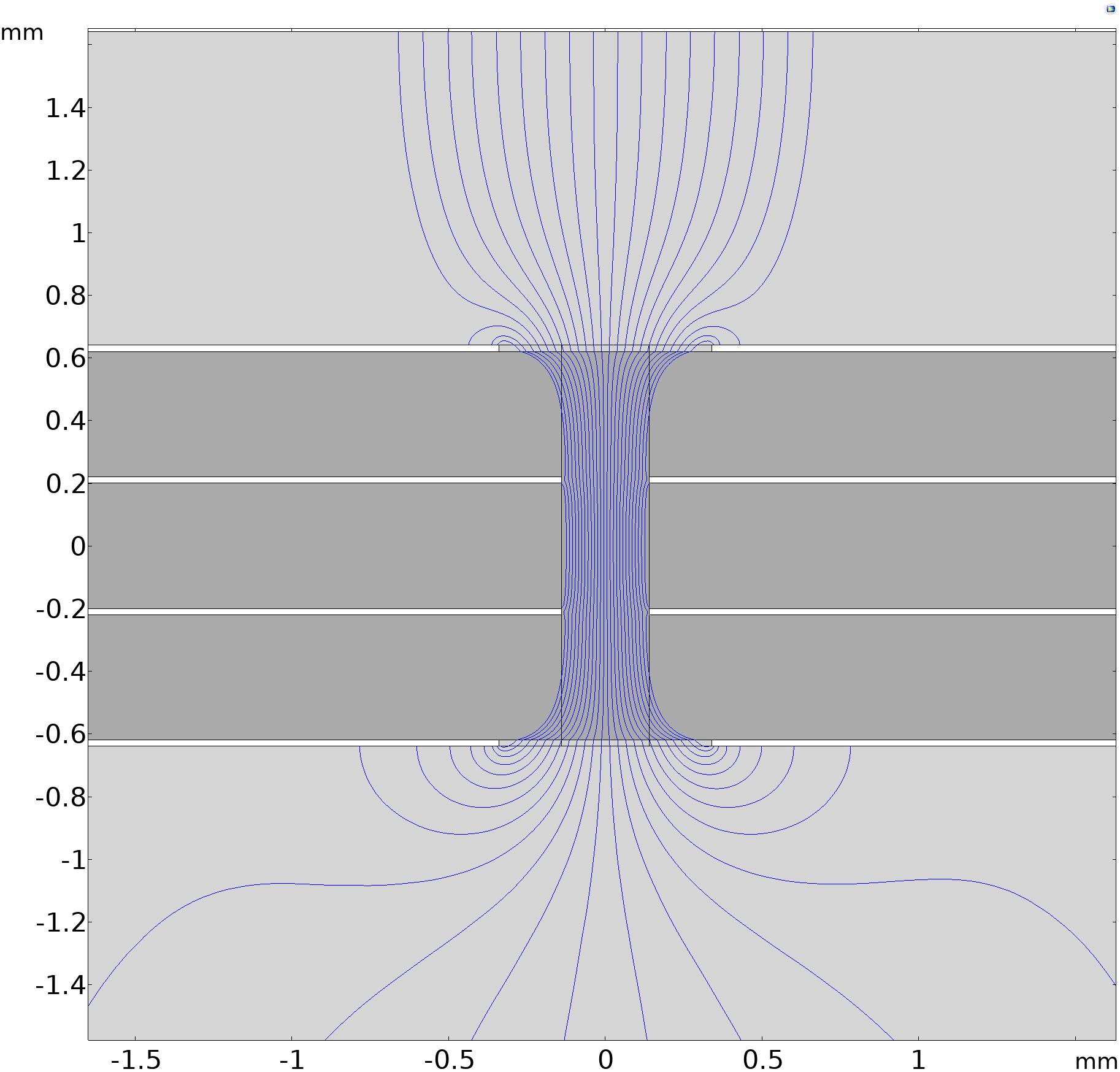}}
		\quad
		 \subfloat[]
		 [\label{fig:COMSOL_FULL1}] {
		 \includegraphics[width=0.5\textwidth]{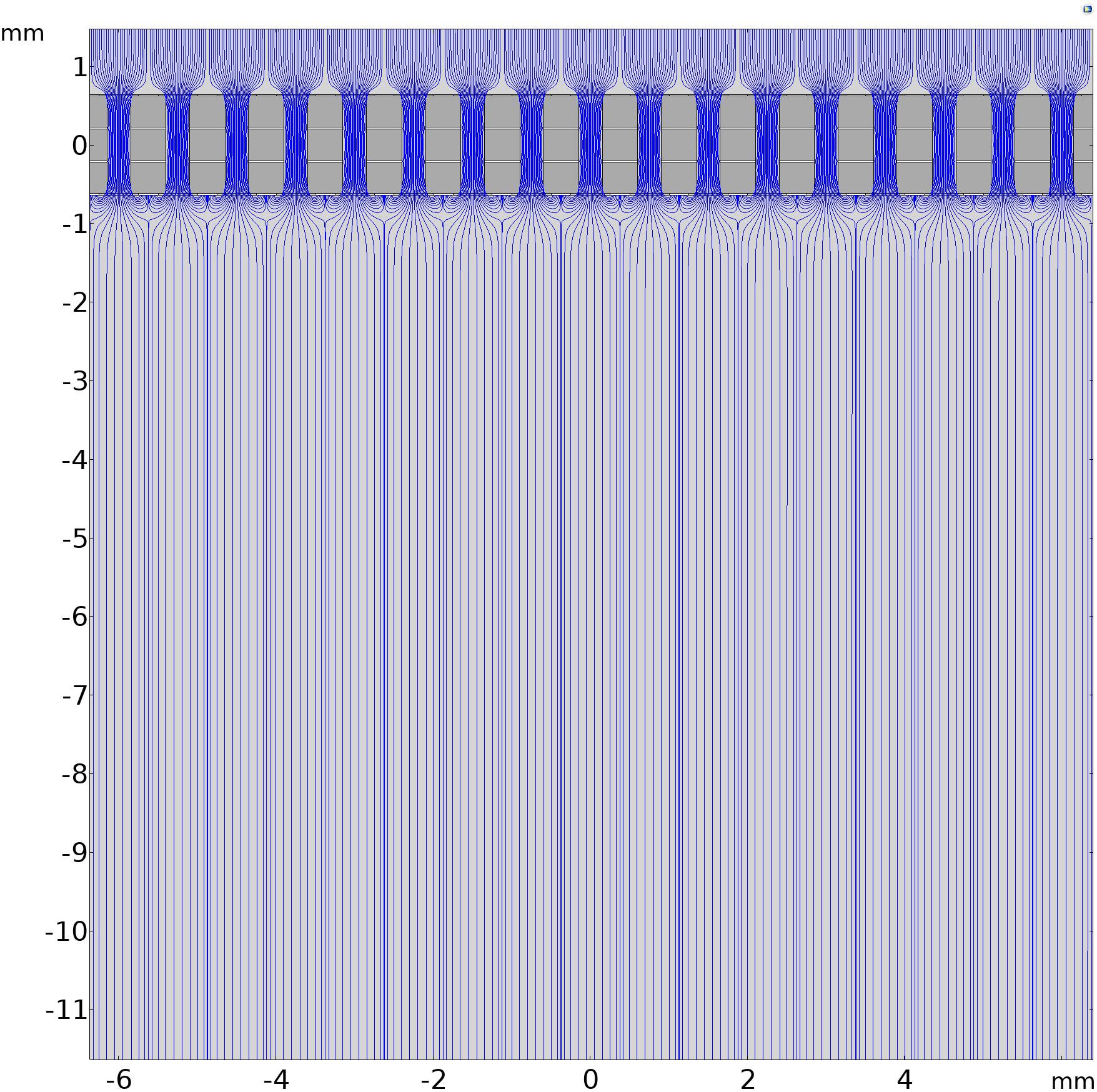}}
		 \subfloat[]
		 [\label{fig:COMSOL_FULL2}] {
		 \includegraphics[width=0.5\textwidth]{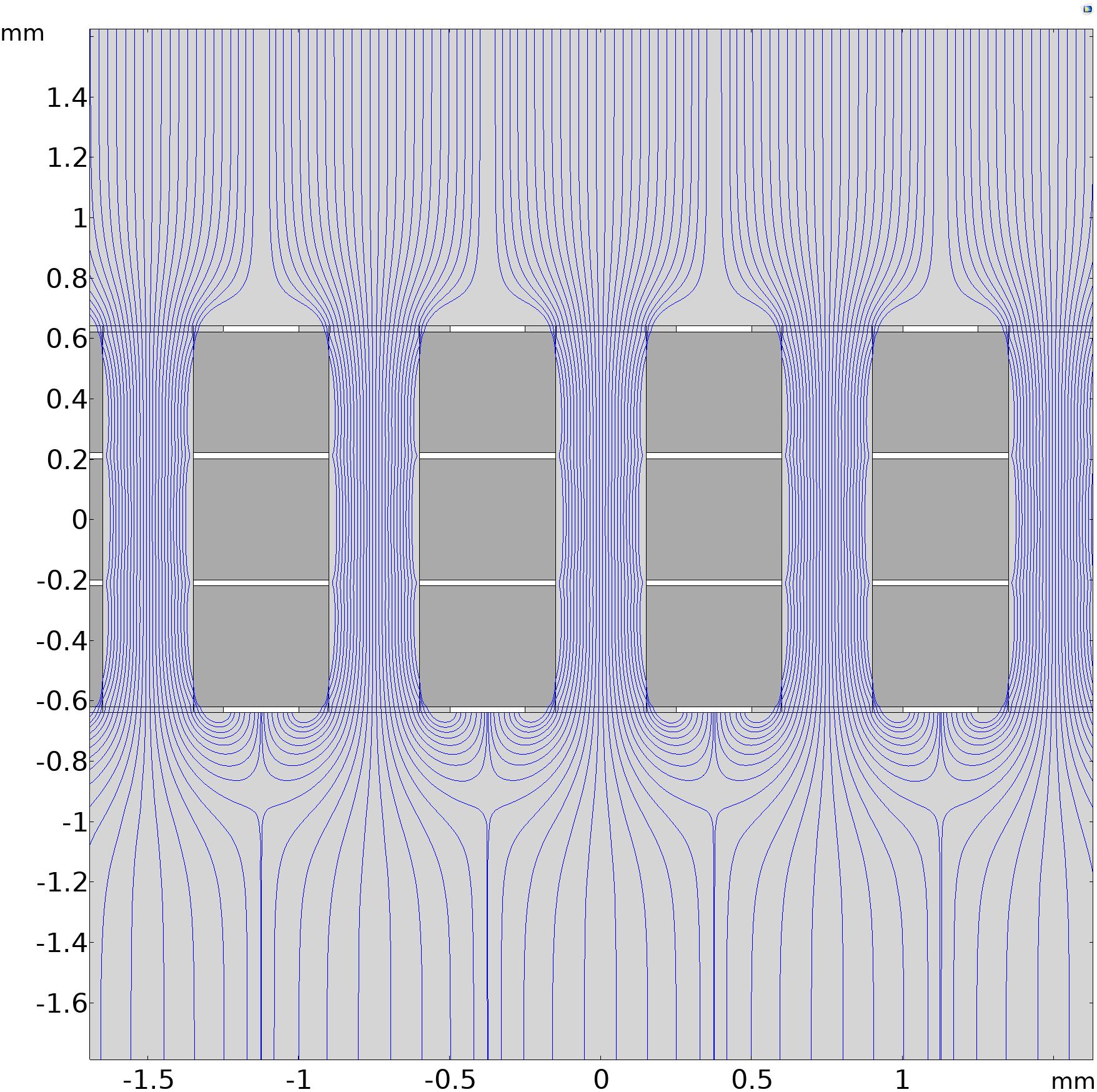}}
	\caption{(Color online) Electric ﬁeld in the region of the holes for ROW a), b) and for FULL c), d) obtained with COMSOL Multiphysics software. The blue field lines start from a plane parallel to the M-THGEM foils placed in the middle of the central hole, while the red field lines in a) originate from the cathode.}
	\label{fig:field_lines}		
\end{figure*}


\paragraph{The induction region}
The induction region, 1 mm thick, is delimited by the M-\thgem{} top electrode and a segmented anodic plate. The avalanche electrons emerging from the holes of the M-THGEM are directed towards the anode by an electric field of about 1000 V/cm.
In the prototype tests here presented, the segmented, position-sensitive anode~\cite{Carbone2012} was replaced by a single anode plate to collect the full avalanche charges in a single pad.\\

In Fig.~\ref{fig:operation_principles} the operating principle of the tracker is illustrated. 
When an incident charged particle crosses the gas in the drift region, it produces a number of positive ion and electron pairs along its track. The primary electrons drift with constant velocity towards the multiplication region as a result of the uniform electric field. 
Then they are funneled inside the holes of the M-\thgem{} where they are multiplied by the gas avalanche process, as a result of the intense electric field established in the  M-\thgem{} holes. The resulting avalanche electrons are then extracted from the holes and collected onto the segmented readout anode.
The signal induced on each element of the anode is sent to a charge preamplifier and then to a digitizer. 
The charges collected on the segmented anode allows the horizontal positions and angle to be determined. The vertical
positions and angle are extracted from the electron drift times.

Two different M-\thgem{} layouts have been tested in the present study: both of them are three-layer M-\thgem{} and have an area of $107 \times 107$~$\mbox{mm}^{2}$, but they show different 
hole patterns, as can be seen in  Fig.~\ref{fig:thgem_holes}.
The first type (ROW M-\thgem{}, \rowthgem{} in the following) has just five rows of holes, laid parallel to the entrance and exit surfaces of the detector, and separated by 18 mm one from the other.
\rowthgem{} was manufactured by Zener s.r.l. and is composed by three layers of PCB (0.40 mm thick), each one coated with copper (0.020 mm thick), resulting in a total thickness of 1.28 mm.\\
In the second type (FULL M-\thgem{}, \fullthgem{} from now on) the holes cover the whole active area following a square pattern (the holes are placed at the vertices of a square as shown in Fig.~\ref{fig:thgem_holes}). 
\fullthgem{} was produced by Shenzhen HeLeeX Analytical Instrument Co. Ltd. and is made up of alternate layers of ceramic SD103K (0.422 mm thick) and copper (0.026 mm thick) for a total thickness of 1.37 mm.

ROW was designed to be coupled with an anode segmented in strips perpendicular to the hole rows as sketched in Fig. 2 and described in Ref.~\cite{TDR2021}. In this case each row of holes of the M-THGEM defines a $z$ coordinate. Just the electrons produced in correspondence of each row are multiplied and can generate a signal in the segmented stripped anode.  If the anodic strip size is sufficiently large, a typical track with an angle $\theta$ on the $x-z$ plane generates signals on five strips, corresponding to the crossing of the five M-THGEM rows with the ion track. The $x$ coordinate is defined by the corresponding hit strip. Thus, this setup based on ROW and stripped anode could allow to track the incident ion without measuring the charge distribution, resulting in a fast and cheap approach for signal transmission. A drawback of this setup is that the $x$ position resolution is limited by the size of the anodic strips. Moreover, ambiguities on the track reconstruction for trajectories parallel to the anodic strips could appear.
On the other hand, a more standard setup based on FULL coupled with an anode segmented in pads distributed in the $x-z$ plane, as the one described in Ref.~\cite{Cappuzzello2021}, overcome the above-mentioned problems, even if the readout of a larger number of channels with respect to the ROW setup is requested.
The main characteristics of the two M-THGEM foils are summarized in Table~\ref{tab:thgem_characteristics}.

The different hole geometries of the two M-THGEM foils have a strong effect in shaping the electric field, especially close to the M-THGEM surface. 
Calculations of the electric field in the detector, using the two kinds of M-THGEM foils, have been performed using the finite-elements software COMSOL Multiphysics~\cite{COMSOL}. 
In Fig.~\ref{fig:field_lines} the field lines for the two cases are shown. 
The electrical configuration of the detector adopted in these calculations is: 800 V potential difference across the drift region, 200 V across each M-\thgem{} layer, and 50 V across the induction region. 
In both cases it is evident that far from the holes the electric field is quite uniform, whilst close to the M-THGEM plate the electric field lines focus inside the holes.
The bending of the field lines, due to the focusing effect, occurs in an area that is much larger for \rowthgem{} compared to \fullthgem{}. 
The loss of the electron collection efficiency in the ROW is actually given by the amount of field lines that end up on the M-THGEM bottom electrode.
This behavior strongly affects the electronic transparency that results smaller for \rowthgem{} and larger for \fullthgem{}, as will be shown in Sect.~\ref{sec:characterization}.

\subsection{The test set-up}

\begin{figure}[t]
    \centering
    \includegraphics[width=0.5\textwidth]{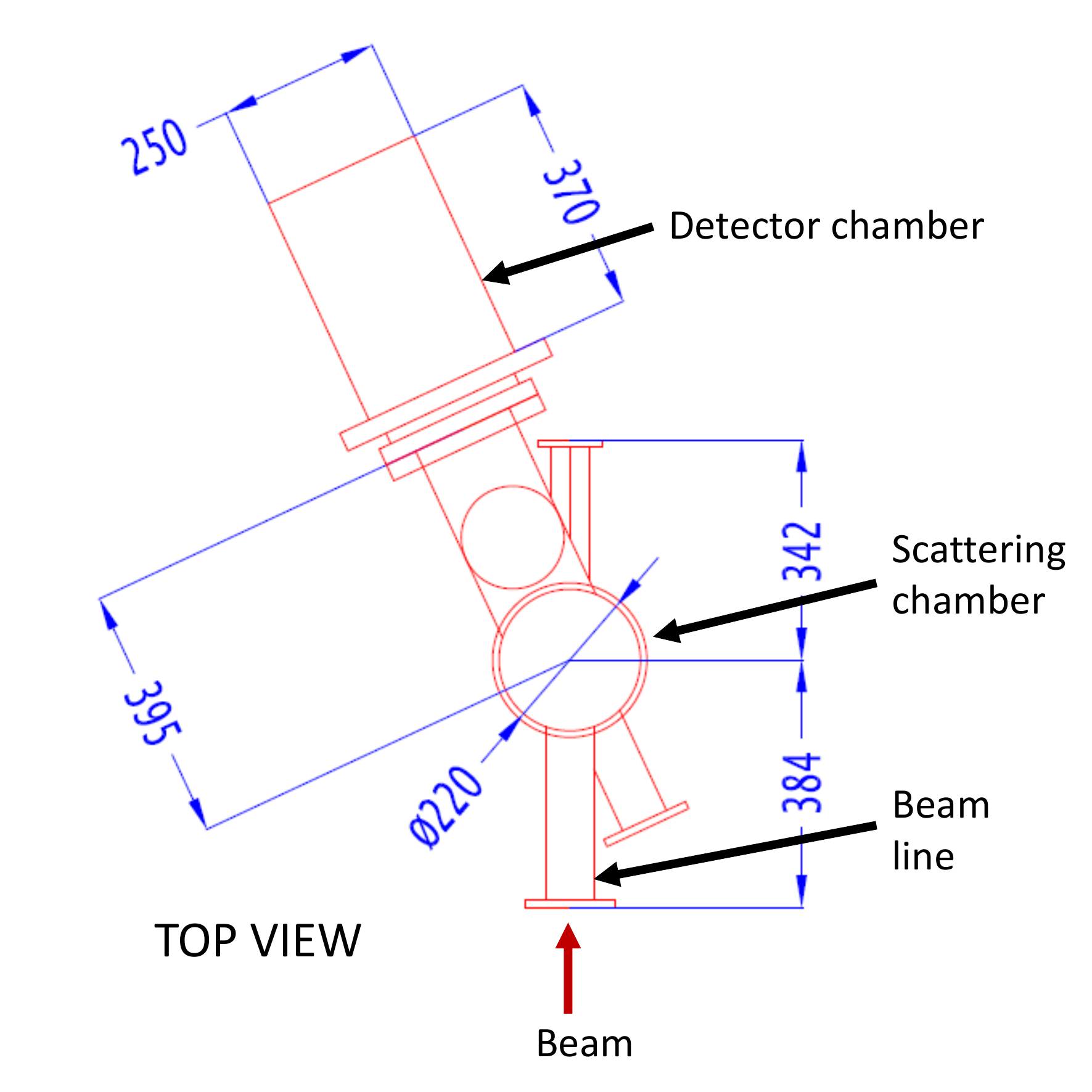}
    \caption{(Color online) Top view drawing of the TEBE facility at INFN-LNS.}
    \label{fig:tebe}
\end{figure}

The tests of the tracker prototype were performed at INFN - LNS (Catania, Italy) at the TEBE (TEst BEnch) facility (Fig.~\ref{fig:tebe}). TEBE is an equipped beam line mainly dedicated to tests and characterization of detectors for the NUMEN project. Two vacuum chambers are arranged along the beam line, the first (scattering chamber) is equipped with a movable target holder that can house many targets. 
The second chamber (detector chamber) can be filled with gas and
is isolated from the scattering chamber by a 2.5~$\mu m$ thick Mylar window. 
The window is thick enough to withstand a pressure difference of more than 100 mbar and thin 
enough to minimize the energy and angular straggling, even for low energy heavy-ion beams.
The detector chamber houses the tracker prototype and is rotated at $\ang{30}$ with respect to the 
beam direction. During normal operation it is filled with 99.95\% pure isobutane at pressure ranging from 10 to 40 mbar. 
A typical gas flow rate of 130 sccm (standard cubic centimeter per minute) is constantly maintained and controlled by a mass flow controller.

In the tests, a $^{241}\mbox{Am}$ radioactive $\alpha$-particle source with a nominal activity of 52~kBq was placed inside the detector chamber, in front of the tracker prototype. The source was  
collimated in order to reduce the incident particle rate on the detector to about 200 Hz. A remote-controlled shutter was placed in front of the source to stop $\alpha$-particles from reaching the detector in between two experimental runs or whenever needed.

An accelerated ion beam was also used 
in some of the runs, in order to explore the response of the detector to heavy-ion beams at different rates.
A $^{18}$O$^{8+}$ beam at 270 MeV was delivered by the LNS Superconducting Cyclotron with intensities, measured
by two Faraday cups located upstream and downstream of the target, ranging from 100 to 900 pA.  Two gold targets with  thicknesses of 
0.97 and 9.6~$\mbox{mg/cm}^{2}$ were used as additional element to change the rate of particles scattered to the detector from a few tens of pps up to over 3 kpps.

A 16 channel high-voltage power supply (CAEN SY5527 mainframe with A1515 board + A1015G adapter~\cite{Caen5527}), 
specifically designed for powering multiple GEM detectors, was used to supply the required voltages.

\begin{figure}[t]
	\centering
		 \includegraphics[width=0.4\textwidth]{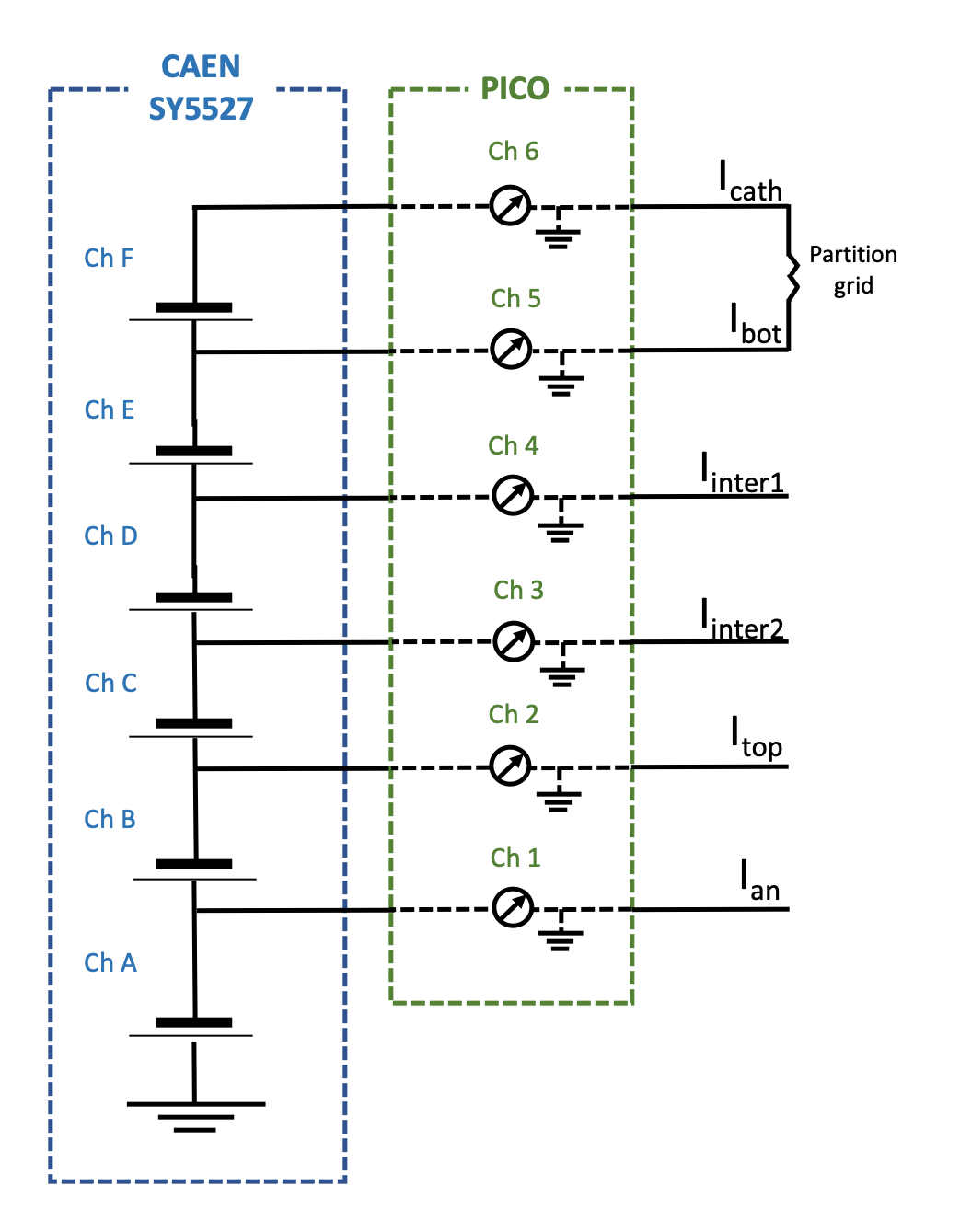}
	\caption{(Color online) Scheme of the biasing and measuring systems based on CAEN SY5527 and PICO.}
	\label{fig:caen_pico}		
\end{figure}

The currents induced on the different electrodes of the tracker were measured by PICO, a high voltage (0 - 1000 V) seven-channel picoammeter, designed and assembled at INFN - Napoli (Italy). PICO was designed to act as fast monitor of the voltage and current of the triple-GEM detector for the CMS experiment \cite{Abbas2021}, but it can be implemented as a general purpose device for other applications with micro-pattern gaseous detectors. The picoammeter hosts 7 ADCs (24bit) and it is capable of measuring voltage with a precision better than 10 mV and current with a precision of about 15 pA for the [$-$16; 4] $\mu$A full-scale range and about 2 nA for the [$-$0.8; 0.2] mA full-scale range.
In particular, in our tests the current on the bottom electrode of the M-\thgem{} (\Ibot{}, see Sect. \ref{sec:characterization}) is typically measured in the low precision scale, while the other currents have values compatible with the high precision one. A detailed scheme of the electrical connection between CAEN SY5527 and PICO is illustrated in Fig. \ref{fig:caen_pico}.

\section{Current-voltage characterization of the gas tracker prototype}\label{sec:characterization}

This section presents the results of the current-voltage characterization and the gain and ion backflow measurements for different voltages and different gas pressures configurations.

\begin{figure}[t]
	\centering
		 \includegraphics[width=0.49\textwidth]{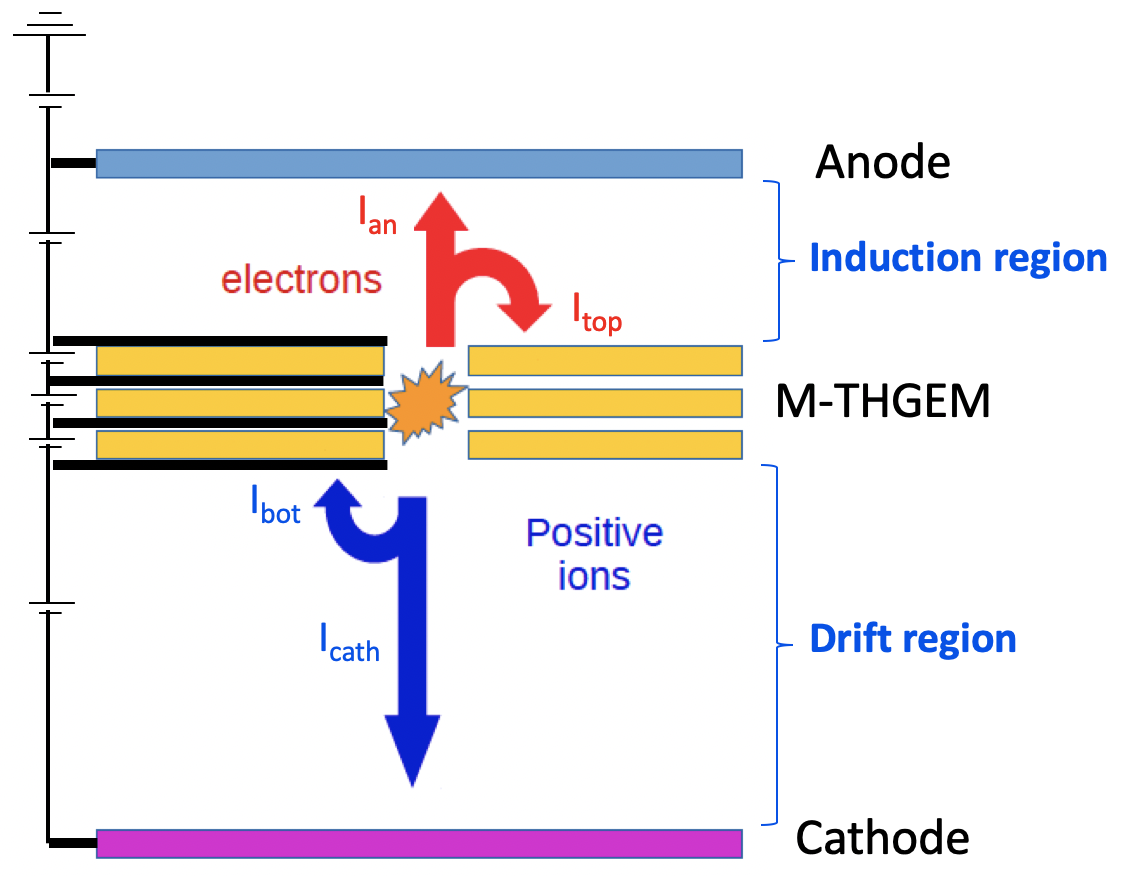}
	\caption{(Color online) Biasing scheme of the detector with an illustration of the measured currents.}
	\label{fig:working_ch}		
\end{figure}

A schematic electrical diagram of the detector is shown in Fig.~\ref{fig:working_ch}. 
The currents flowing through the detector electrodes are:
\begin{itemize}
    \item \Ian{}, fraction of the avalanche electrons that are transferred to the readout anode and contribute to the detected signals;
    \item \Itop{}, fraction of the avalanche electrons and ions (see Sect.~\ref{sec:current_vind}) collected onto the M-THGEM top electrode;
    \item \Ibot{}, fraction of the avalanche ions collected onto the M-THGEM bottom electrode;
    \item \Icath{}, fraction of the avalanche ions that flow back to the cathode through the drift region.
\end{itemize}

When no beam or $\alpha$-particles cross the detector, the dark current measured by PICO is the parasitic current due to the ohmic resistance of the M-THGEM insulator substrate, sandwiched between the electrodes at different potential.
Due to the partition grid, a typical current of few $\mu$A is measured by the cathode (\Icath{}) and bottom (\Ibot{}) electrodes.
The currents collected by the intermediate M-THGEM electrodes (not indicated in 
Fig.~\ref{fig:working_ch}) are in most of the cases below the precision of the picoammeter and will not be discussed in the following.
In the present tests the anode was biased at $-20$~V.

\begin{figure*}[htbp]
    \centering
    \includegraphics[scale=0.7]{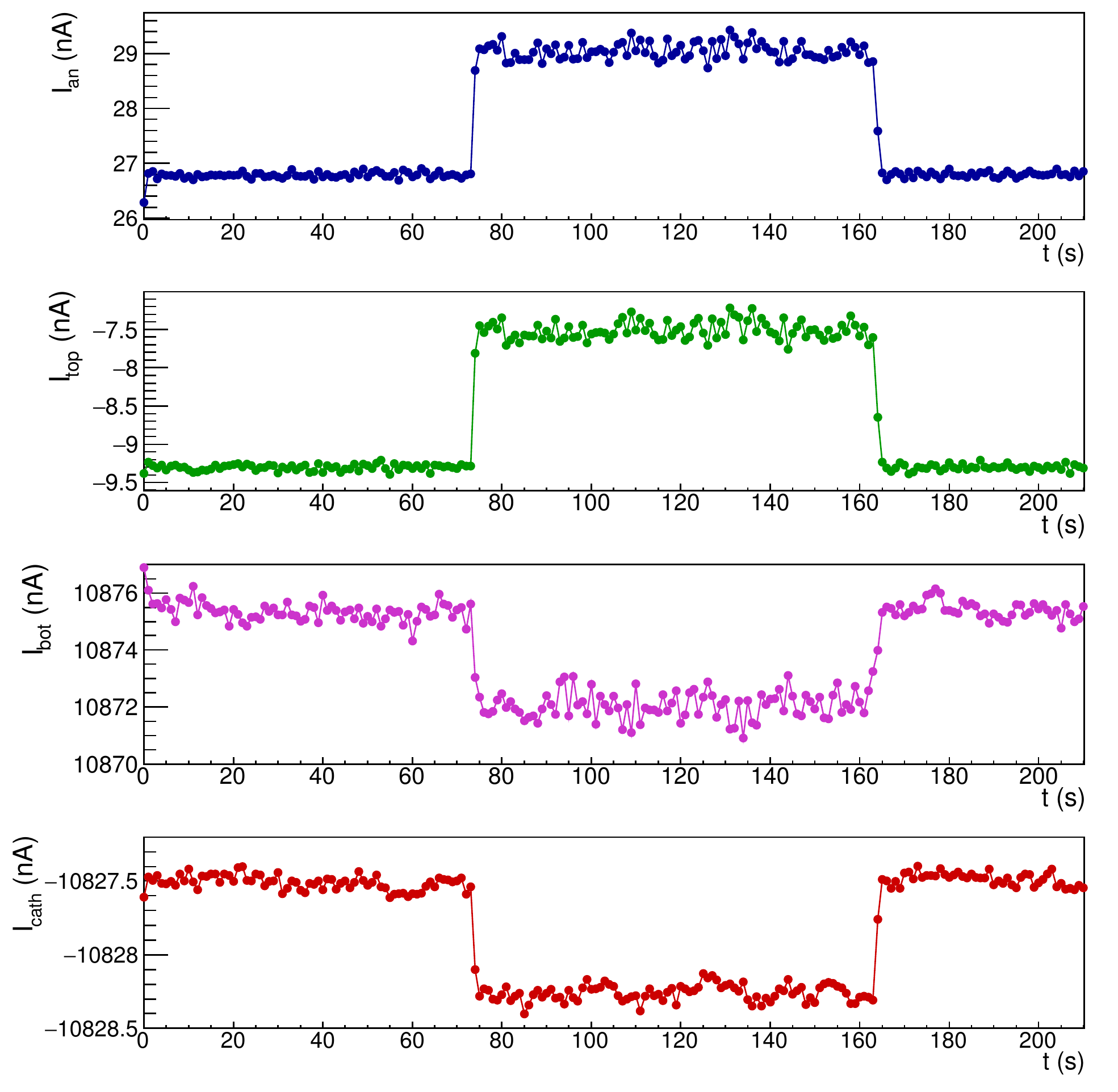} 
    \caption{(Color online) Measured currents as a function of time in a typical run. From the top to the bottom of the figure, the current of anode, top layer of the M-THGEM, bottom layer of the M-THGEM and cathode electrodes are shown. The two abrupt changes in the 
    current at $\approx$ 80 and $\approx$ 160~s correspond to the opening and closure of the shutter in front the 
    $\alpha$-particle source, respectively. The measurement was performed with $P = 30 $ mbar, \Vind{}$ = 40 $ V, \Vthgem{}$ = 200$ V, and \Vdrift{}$ = 1000$ V.}
    \label{fig:current_file}
\end{figure*}

In the measurements with $\alpha$-particle source or $^{18}$O beam, for each electrical configuration, a run of about 200~s was performed. During the first 80~s the shutter in front of the source was closed, in the next 100~s it was opened and finally for the last 20~s it was closed again. 
The net current induced on each electrode was extracted as the difference between the average current measured with closed and open shutter.
The error on the net current was obtained in the following way. First, we calculated the error on the average current with closed shutter ($\Delta I_{closed}$) as the quadratic sum of the statistical contribution and the precision of the picoammeter. The same procedure was applied to deduce the error on the average current with open shutter ($\Delta I_{open}$).
Then, the total error assigned to each point is the quadratic sum of $\Delta I_{closed}$ and $\Delta I_{open}$.
This method allows to measure currents due only to the charged particles crossing the detector; any contribution from dark currents circulating in the detector and/or possible bias in the picoammeter current measurements are removed.
Possible sources of systematic errors are electronic noise, change in temperature and pressure and stability of the bias supply. During the runs the gas temperature and the pressure were within 1$^\circ$C and 0.5 mbar, respectively. We estimate the total systematic error by comparing the measurements performed in the same experimental conditions as a maximum of 10\%.
An example of the currents measured in a single run as a function of time is shown in Fig.~\ref{fig:current_file}. In the explored experimental conditions we observe a good long-term gain stability, which is a sign of
negligible charging up of the insulator substrate.\\

The currents have been measured, varying four parameters that affect the behavior of 
the detector, namely: 
\begin{itemize}
    \item the gas pressure;
    \item the voltage difference applied to the induction region (\Vind{});
    \item the voltage difference applied across a single \thgem{} layer (\Vthgem{}), which was the same for all the three layers of the M-\thgem{} except when explicitly mentioned;
    \item the voltage difference applied to the drift region (\Vdrift{}).
\end{itemize}
Each current-voltage characterization was obtained changing
only one parameter at a time and 
keeping fixed all the others. 
In the figures shown in this section, the error bar of the measured points is always included, if not visible it is because it is smaller than the marker size.


\subsection{Current-\Vind{} characterization} \label{sec:current_vind}

\begin{figure}[t]
    \centering
    \includegraphics[scale=0.4]{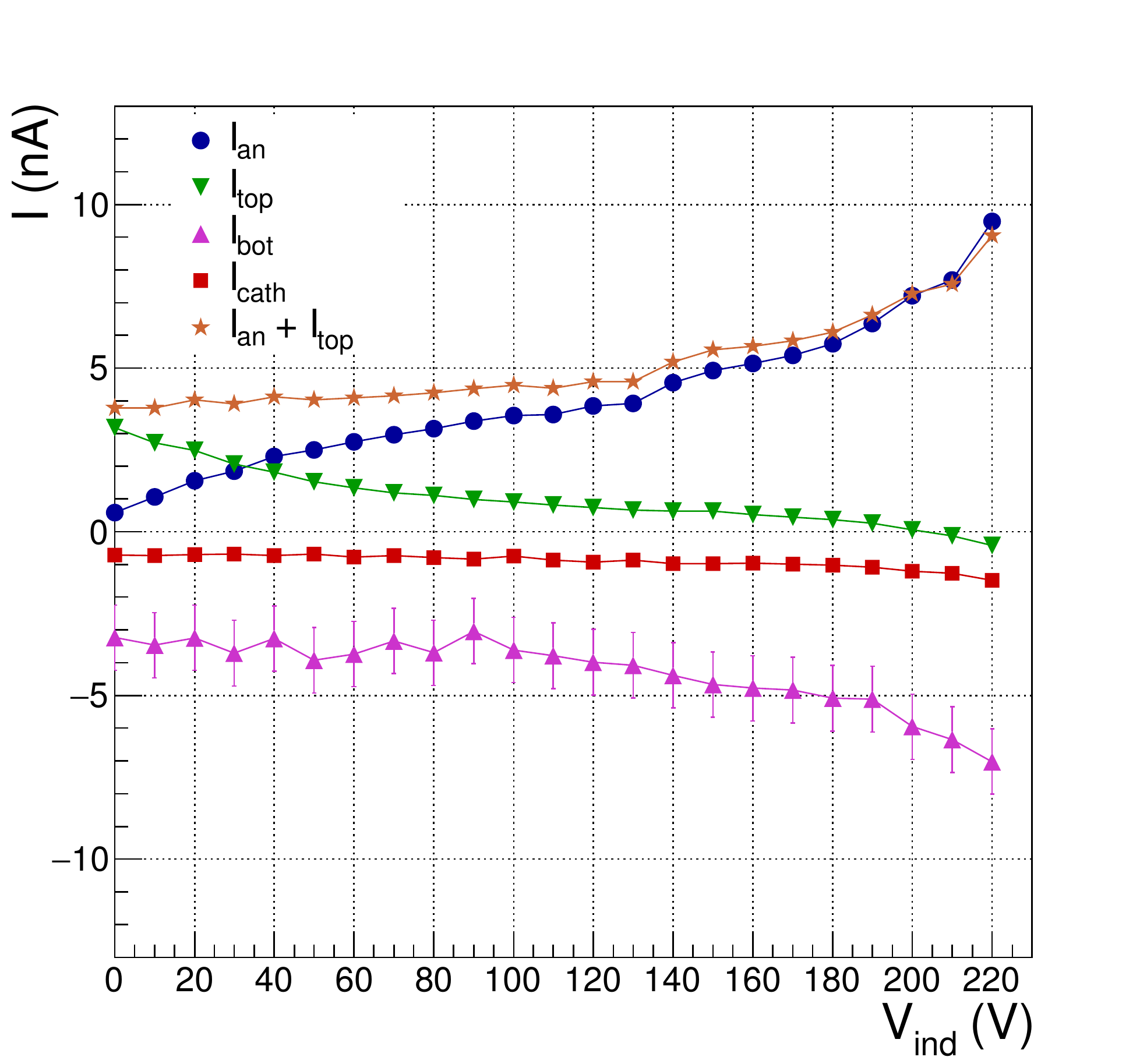}
    \caption{(Color online) Current-\Vind{}  characterization for \fullthgem{} at $P=$ 30~mbar, \Vthgem{} $=$ 220~V and 
    \Vdrift{} $=$ 1000~V.
    }
    \label{fig:induction_FULLTHGEM_30mbar}
\end{figure}

The \Vind{} voltage determines the electric field in the induction region, therefore it impacts on the efficiency of extracting the avalanche electrons from the M-THGEM holes and collecting them onto the readout anode.

The characterization of the measured currents as a function of \Vind{} was performed by fixing the gas pressure, \Vthgem{} and 
\Vdrift{} and changing \Vind{} in steps of 10 or 20~V from 0~V up to the discharge value. 
Different configurations of \Vthgem{}, \Vdrift{} and pressure were used, as listed in
Table~\ref{tab:ROWTHGEM_vind}.

As an example, in Fig.~\ref{fig:induction_FULLTHGEM_30mbar} the characterization of V$_{ind}$ for P = 30 mbar, \Vthgem{}~$= 220$~V and \Vdrift{}~$= 1000$~V for \fullthgem{} is shown. The main feature 
of the plot is that as \Vind{} increases, the magnitude of \Ian{} increases, whilst \Itop{} 
decreases. Therefore the value of \Vind{} modulates the ratio $I_{an}/I_{top}$.
The sum of the currents read on top of the M-THGEM and anode electrodes (\Itop~+~\Ian), 
that is the total negative charge (TNC) produced by the M-\thgem{}, is also shown in the figure. 
Up to about 140~V, the TNC is approximately constant, while for values larger than 140~V it starts to increase. 
This indicates that, up to 140~V, the stronger the electric field in the induction region, the larger the fraction of secondary electrons collected by the anode and the lower the fraction of the electrons hitting the top electrode. 
For \Vind{} larger than 140~V, the rise of the TNC can be explained considering that the electric field in the induction region becomes strong enough to produce charge multiplication. It can be estimated that the maximum gain obtained in the induction region in these electrical conditions is of a factor 2.
For higher values of \Vind{}, \Itop{} becomes negative. 
In this condition the amount of the positive ions produced in the induction region and collected by the top of the M-\thgem{} is larger than the electrons collected by the same electrode.
For the \fullthgem{} with this electrical and pressure configuration an operational value of about \Vind{}~$= 120$~V allows that a large fraction of the electrons is collected on
the anode and the multiplication is mainly confined inside the M-\thgem{} channels.

A similar behavior is observed for the other electrical configurations and pressures and for the \rowthgem{}.


\subsection{Current-\Vdrift{} characterization}

\begin{figure}[h!]
    \centering
    \includegraphics[scale=0.53]{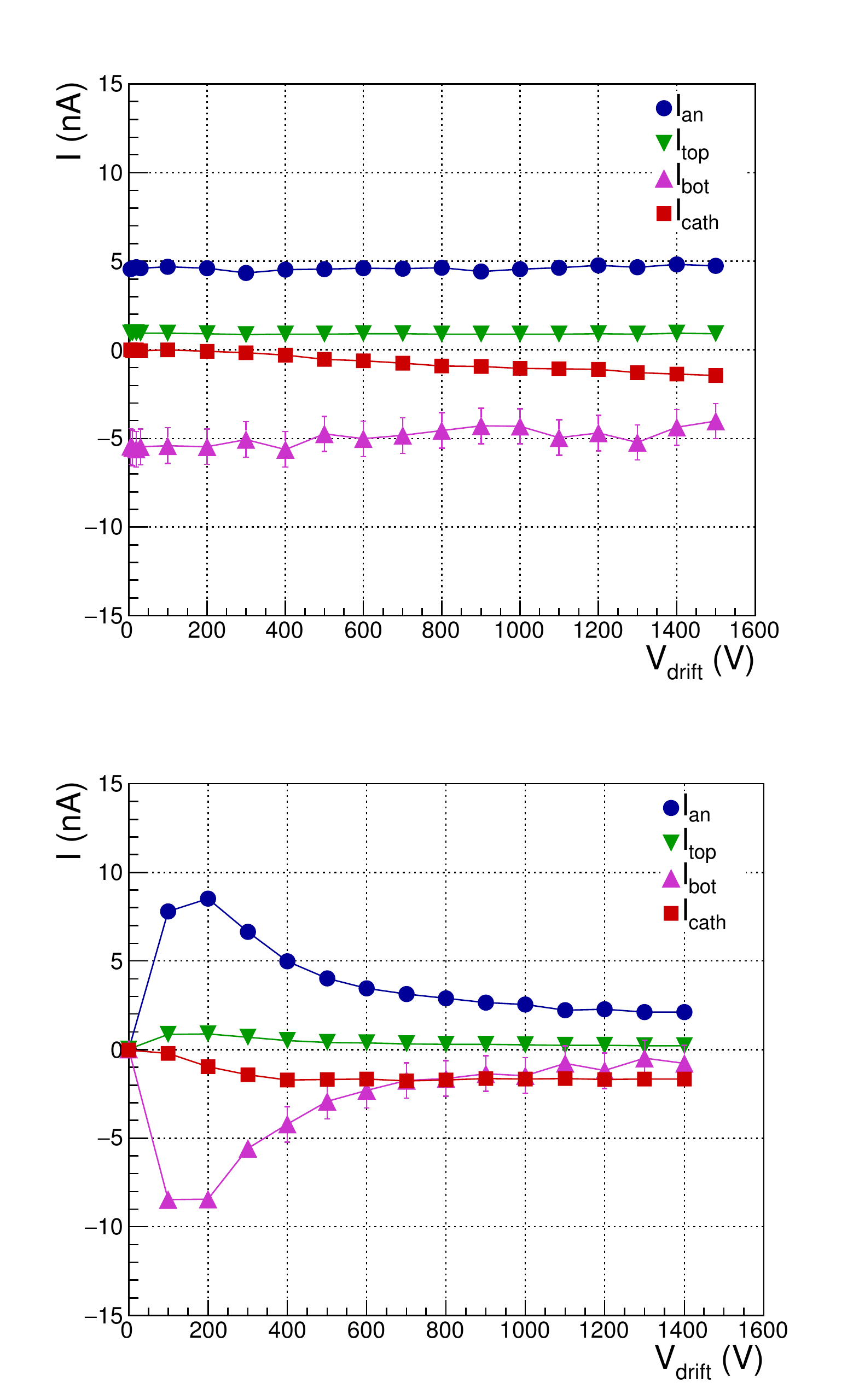} 
    \caption{(Color online) Current-\Vdrift{} characterization for \fullthgem{}
    at $P = 30$~mbar, \Vind{} $= 70$~V and \Vthgem{} $= 240$~V (top) and for \rowthgem{} at $P = 30$~mbar, \Vind{} $= 120$~V and \Vthgem{} $= 220$~V (bottom).
    }
       \label{fig:drift_ROWFULLTHGEM_30mbar}
\end{figure}

The study of the \Vdrift{} current-voltage characterization was performed varying the \Vdrift{} value from 0~V up to the discharge value. A summary of the different configurations explored 
during the tests is given in Table~\ref{tab:ROWTHGEM_vdrift}.

For the \fullthgem{} the anodic current is almost completely flat and no variation with 
\Vdrift{} is evident for all the configurations explored. An example of the behavior is shown in Fig.~\ref{fig:drift_ROWFULLTHGEM_30mbar} (top).

The \rowthgem{} behaves in a very different way. A typical plot is shown in Fig.~\ref{fig:drift_ROWFULLTHGEM_30mbar} (bottom). 
As \Vdrift{} increases the anodic current increases reaching a maximum, 
after this value the current decreases.
For small \Vdrift{} values (less than $\approx$100 V), most of the primary electrons are lost because of recombination effects. Increasing \Vdrift{}, a larger number of primary electrons reaches the multiplication stage resulting in an increment of the anodic current.
Assuming the same electric field in the M-THGEM hole, a larger drift velocity results in a large defocusing of the electrons. Thus, more field lines ends up on the M-THGEM bottom surface, with a decrease of collection efficiency.
The behavior of the anodic current for higher \Vdrift{} values has also been observed in other GEM-based detectors \cite{BACHMANN1999376,Bellazzini1998}.

The different behavior of the two kinds of M-THGEMs has important consequences on the 
operation of the detector. 
In fact the strong dependence of the anodic current with \Vdrift{} makes the gain of \rowthgem{} very sensitive to any variation of \Vdrift{}, on the contrary the behavior of \fullthgem{} is not affected by change of \Vdrift{} in a wide range of values.


\subsection{Current-\Vthgem{} characterization}

\begin{figure}[htbp]
    \centering
    \includegraphics[scale=0.4]{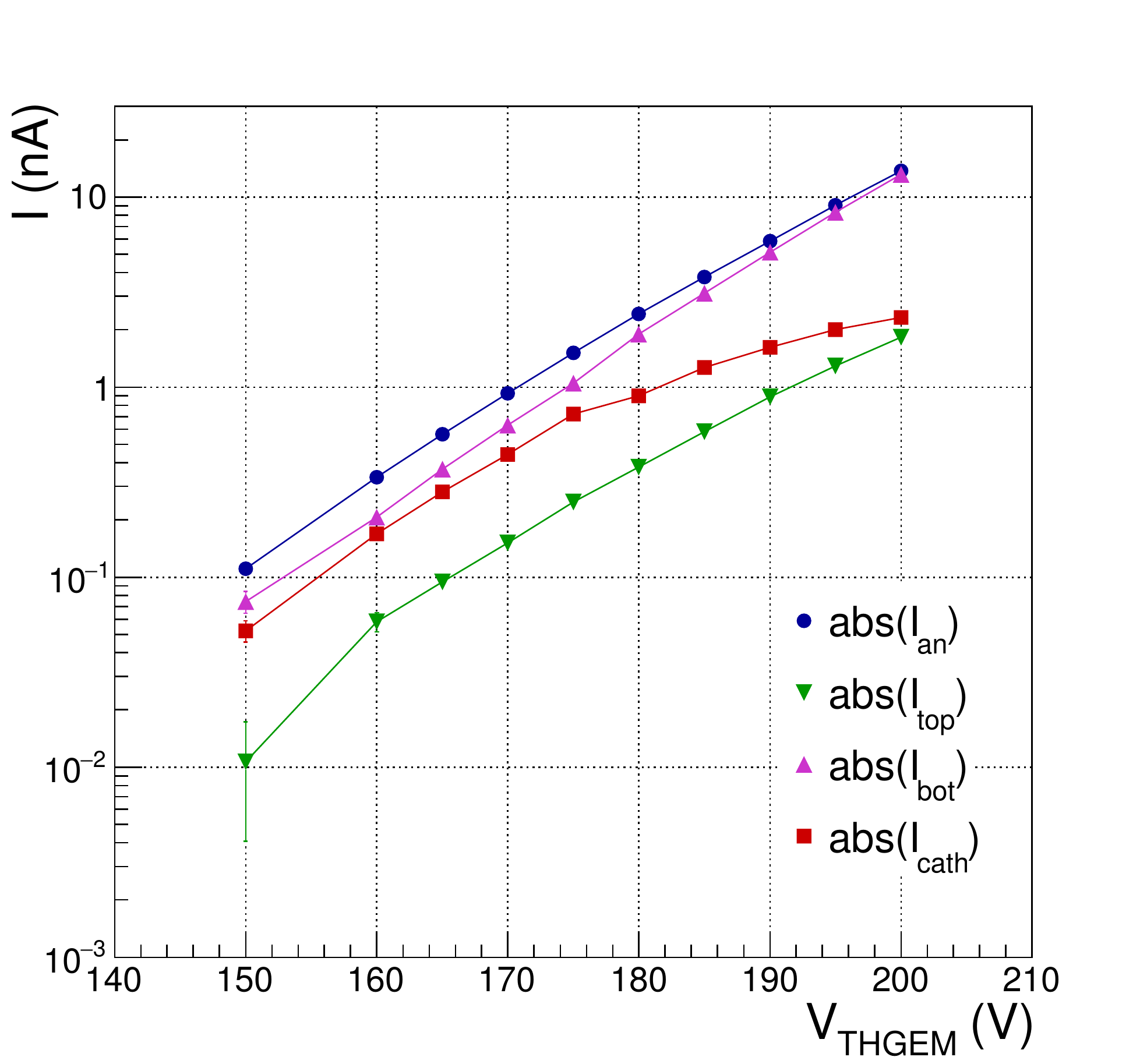}
    \caption{(Color online) Current-\Vthgem{} characterization for \rowthgem{} at $P =$ 10~mbar, 
    \Vind{} $=$ 50~V, and \Vdrift{} $=$ 200~V. The absolute values of the  measured currents are plotted.
    }
    \label{fig:thgem_ROWTHGEM_10mbar}
\end{figure}

Similarly to the previous cases, the \Vthgem{} current-voltage characterization was studied at fixed values of gas pressure, \Vind{}, \Vdrift{} and changing \Vthgem{} at steps of 5 or 10~V from the minimum value for which currents are measurable up to the discharge value.
A list of the different configurations explored in the tests is given 
in Table~\ref{tab:ROWTHGEM_vthgem}. 
As an example, the curve for the \rowthgem{} at a pressure of $P=$~10~mbar is shown in Fig.~\ref{fig:thgem_ROWTHGEM_10mbar}. 
The measured currents increase exponentially with \Vthgem{}, as expected.
For \Vthgem{} larger than 180~V, one observes for \Icath{} and \Ibot{} a deviation from the exponential behavior due to a different charge sharing between cathode and bottom of the M-THGEM. 
A possible explanation of the slight deviation of \Ian{} from the exponential behavior is the gas gain saturation inside the holes.
A similar behavior is observed both for the \rowthgem{} and \fullthgem{} in all the configurations studied.

\subsection{Gain}

The gain of the M-THGEM-based gas detector is here defined as the ratio between the sum of the measured $I_{an}$ and $I_{top}$ currents and the charge of primary electrons per unit of time ($I_{e}$):
\begin{equation*}
    G = \frac{I_{an}+I_{top}}{I_{e}} 
\end{equation*}
$I_e$ was estimated from the number of $\alpha$-particles emitted by the source that enter the detector or the number of scattered $^{18}$O ions reaching the detector per unit of time, and  calculating the energy loss of the particles inside the active volume of the detector using the LISE++ tool \cite{TARASOV2016185}.
Dividing the energy loss ($\Delta E$) by the isobutane mean ionization energy (23~eV), the number of primary electron-ion pairs ($N_{prim}$) can be deduced.
The obtained values of $\Delta E$ and $N_{prim}$ are listed in Table~\ref{tab:energy_loss} for different pressures and for both $\alpha$-particle and $^{18}$O ion.

\begin{table} [htbp]
   \begin{center}
      \caption{Values of energy loss ($\Delta E$) evaluated with LISE++ and corresponding number of primary electron-ion pairs ($N_{prim}$) for both $\alpha$-particle and $^{18}$O ion at different gas pressures ($P$).}
      \begin{tabular}{ccccccc}
		Ion & &  $P$   & & $\Delta E$ & & $N_{prim}$ \\ 
            & & (mbar) & &    (keV)   & & \\
        \hline 
        $\alpha$ & & 10 & & 314 & & 13650 \\
        $\alpha$ & & 20 & & 646 & & 28100 \\
        $\alpha$ & & 30 & & 1000 & & 43500 \\
        $^{18}$O & & 20 & & 1185 & & 51500 \\
        \hline
      \end{tabular} 
    \label{tab:energy_loss}
   \end{center}
\end{table}

As a general observation, the \rowthgem{} has a lower signal to the readout compared to the \fullthgem{} because of its lower electron collection efficiency due to the hole layout, as discussed in Sect.~\ref{gas_tracker_prototype}. 

\begin{figure}[h!]
	\centering
	\includegraphics[scale=0.4]{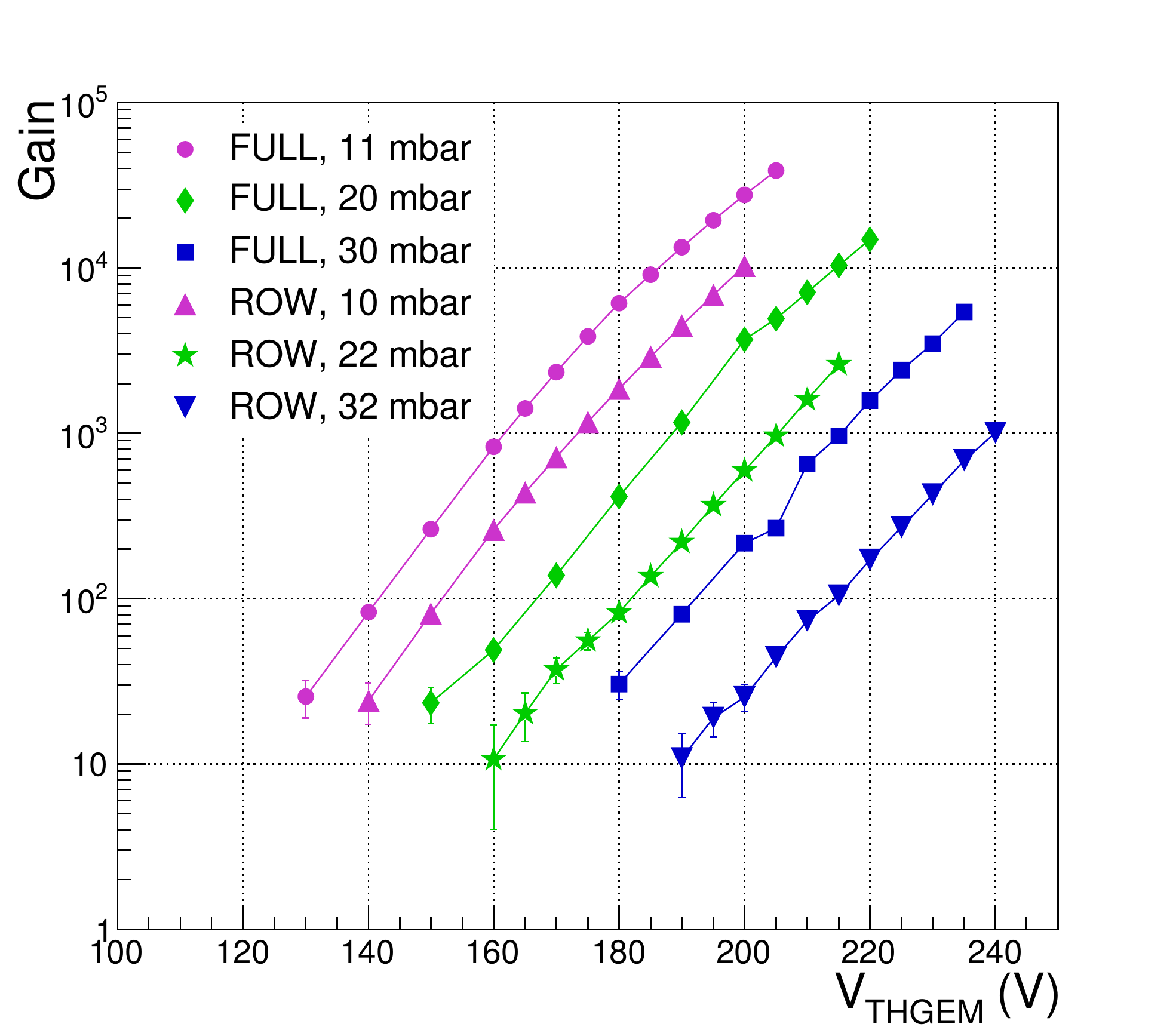}
	\caption{(Color online) Gain obtained as a function of \Vthgem{} for different configurations listed in Table~\ref{tab:ROWTHGEM_vthgem}.}
	\label{fig:gain}
\end{figure}

The gain as a function of \Vthgem{} for different pressures for both \rowthgem{} and \fullthgem{} is shown in Fig.~\ref{fig:gain}. The maximum achievable gain was defined by the onset of the  discharges.
One can see that, as expected, the measured gain exponentially increases as a function of $V_{THGEM}$ for all the explored electrical configurations. A high electron multiplication was achieved also at very low pressure of 11 mbar.
A maximum gain value of 4$\times 10^4$ is reached at 11 mbar. These values are in agreement with those measured for similar detectors ~\cite{Cortesi2020,Cortesi2017,Cortesi2018}.
The decrease of the maximum achievable gain with pressure can be 
explained in terms of discharge when reaching Raether's limit. In fact, this latter defines the maximum number of electrons in a single avalanche (about 10$^7$ - 10$^8$). Thus, if the pressure increases, the number of primary electrons rises as well and the Reather's limit is reached at a lower gain.

\begin{figure}[h!]
	\centering
	\includegraphics[scale=0.4]{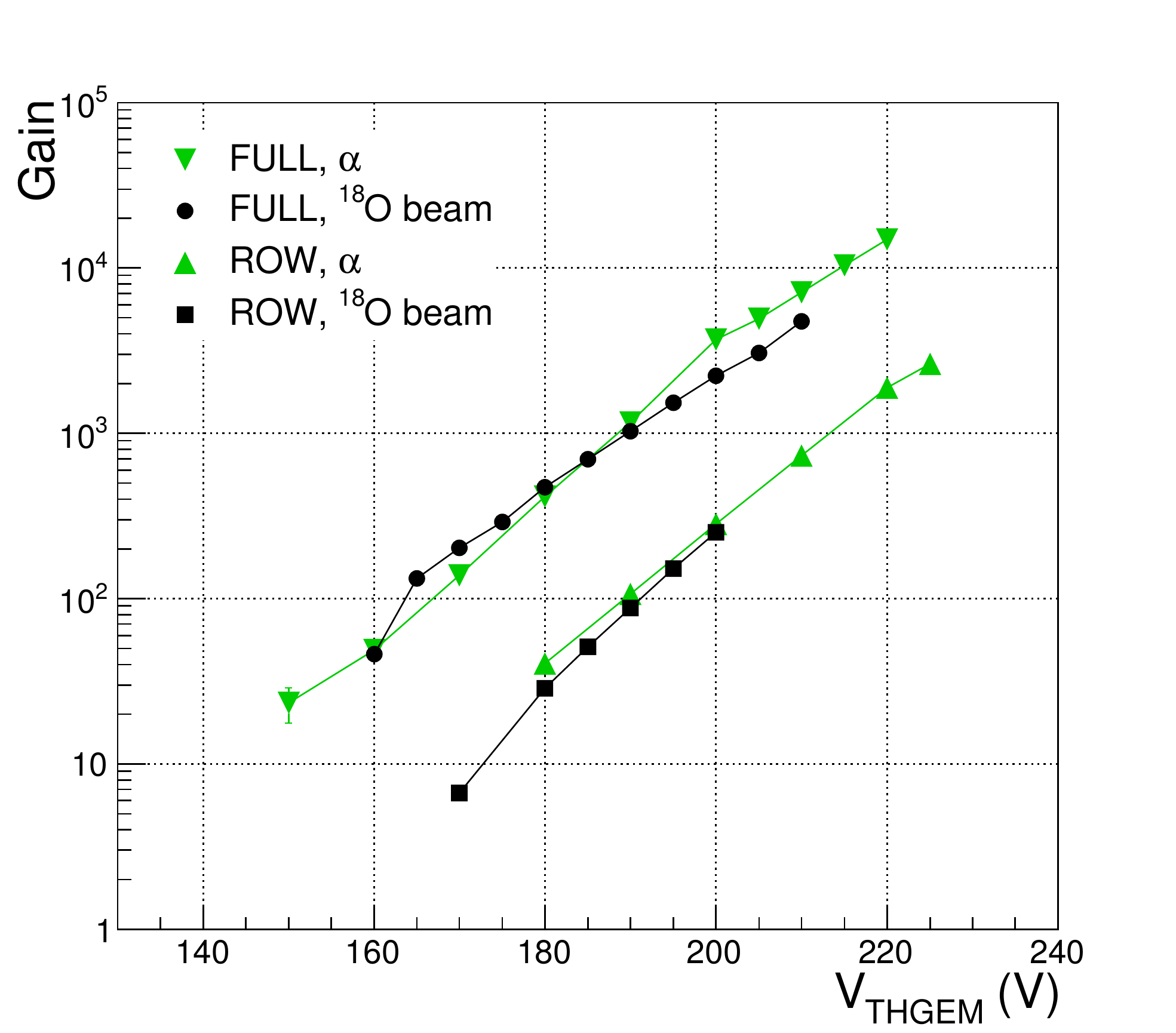}
	\caption{(Color online) Gain for \rowthgem{} and \fullthgem{} as a function of \Vthgem{} for $^{18}$O beam data (black) and $\alpha$-particle source data (green) at $P = 20$ mbar. See
Table~\ref{tab:ROWTHGEM_vthgem} for electrical configurations.}
	\label{fig:gain_beam}
\end{figure}

In Fig.~\ref{fig:gain_beam} a comparison between gains obtained with the $\alpha$-particle source and the $^{18}$O beam for both the M-THGEM types is shown.
There is no significant difference between the gains obtained for the two particle species.


\subsection{Ion backflow} \label{sec:IBF}

\begin{figure}[h!]
	\centering
	\includegraphics[width=0.47\textwidth]{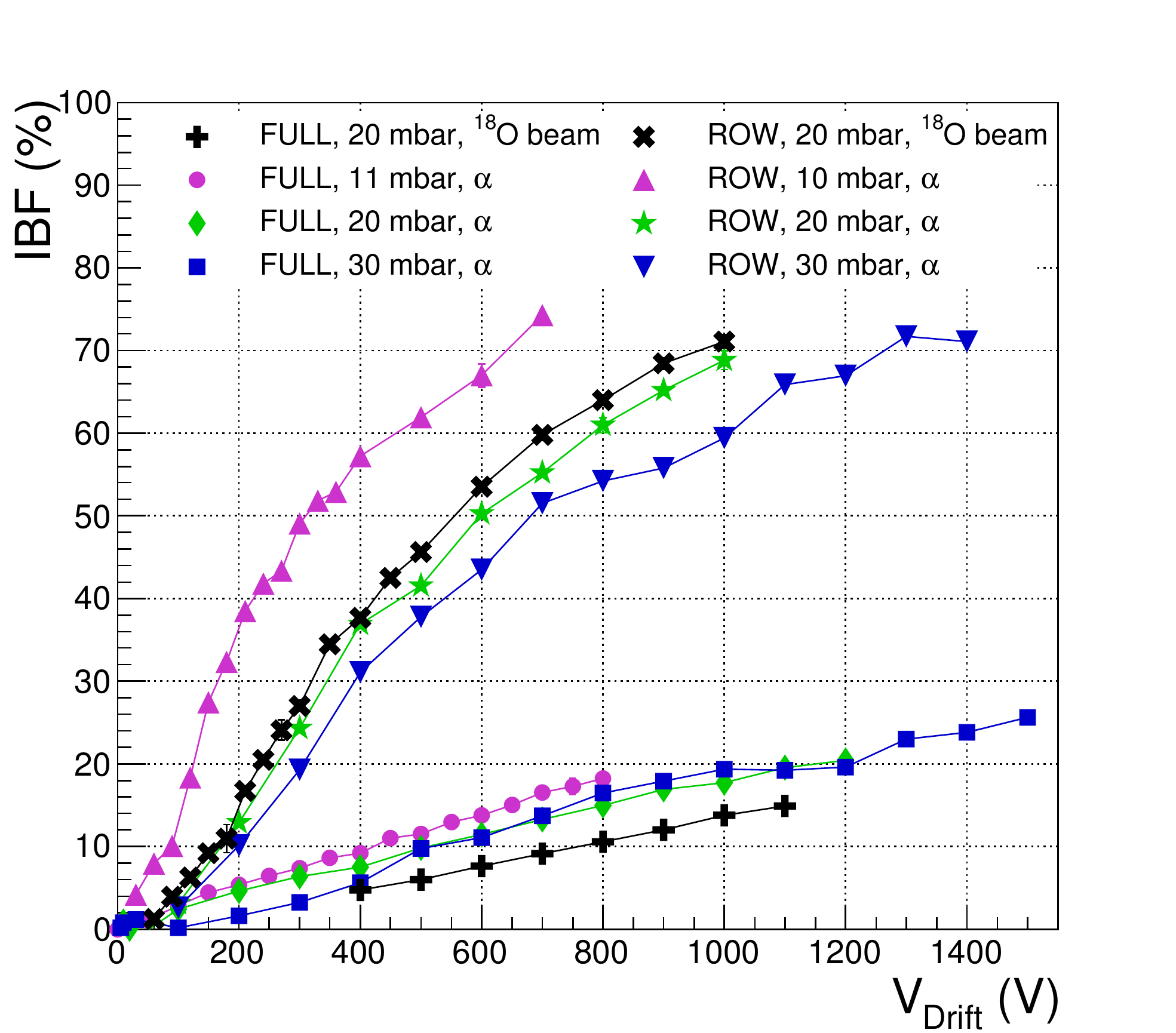}
	\caption{(Color online) IBF for the \rowthgem{} and \fullthgem{} as a function of 
	\Vdrift{} for different gas pressures, using $^{18}$O beam and $\alpha$-particle source. See Table~\ref{tab:ROWTHGEM_vdrift} for electrical configurations.}
	\label{fig:ion_backflow_FULLandROW_alpha_beam}
\end{figure}

\begin{figure}[htbp]
    \centering
    \includegraphics[scale=0.39]{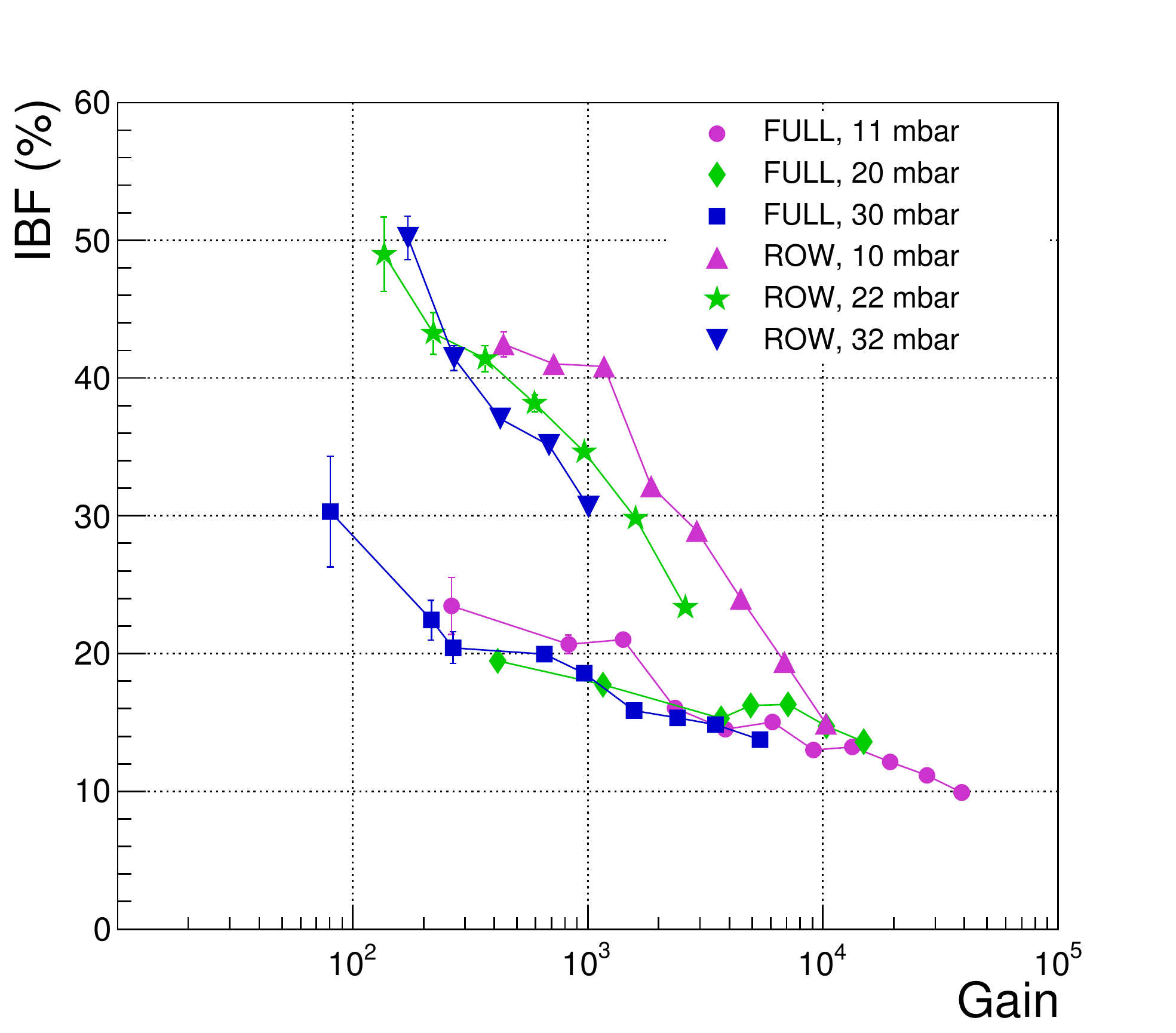}
    \caption{(Color online) IBF for different gas pressures as a function of the gain for $\alpha$-particle source data. See Table~\ref{tab:ROWTHGEM_vthgem} for electrical configurations.}
    \label{fig:IBF_ROW_THGEM}
\end{figure}

The positive ions emerging from the multiplication region are mainly collected on the bottom surface of the M-THGEM whilst a part of them drift back to the cathode. The presence of such a spatial charge in the drift volume can induce severe distortions of the electric field in the drift region that may affect the position resolution and the drift time of the electrons \cite{Sauli2003803,SAULI2006269}.

The Ion BackFlow (IBF) is the fraction of positive ions reaching the cathode and it is usually defined as the ratio between the current at the cathode ($I_{cath}$), and the sum of \Icath{} and \Ibot{}. In our case \Ibot{} is measured with a lower precision than the other currents (see Sect. \ref{sec:exp_set-up}), therefore we decided to use an alternative definition of IBF as the ratio between $I_{cath}$ and sum of negative currents \Ian{} + \Itop:
\begin{equation*}
    IBF = \left|\frac{ I_{cath}}{I_{an}+I_{top} } \right|
\end{equation*}
This definition relies on the fact that the total positive charge (\Icath{} + \Ibot{}) must be equal to the total negative charge  (\Ian{} + \Itop{}) and on the experimental check that the relation: \Icath{} + \Ibot{} = \Ian{} + \Itop{} holds within the experimental error.

In Fig.~\ref{fig:ion_backflow_FULLandROW_alpha_beam} the behavior of the IBF as a function of \Vdrift{} for \fullthgem{} and \rowthgem{} using $\alpha$-particle source and $^{18}$O ion beam is shown. As expected, the IBF presents an increasing monotonic behavior with \Vdrift{}. The \rowthgem{} detector systematically shows a larger IBF than the \fullthgem{} one. In particular, the \fullthgem{} detector reaches a maximum value of about 25\%, to be compared with a maximum IBF of about 75\% for the \rowthgem{} one. This means that for electronic signals on the anode of same amplitude the \rowthgem{} suffers from a larger IBF. 

In Fig.~\ref{fig:IBF_ROW_THGEM} the IBF as a function of the gain for \fullthgem{} and \rowthgem{} is compared. 
The main feature is that IBF has a monotonic trend that decreases with increasing gain. Such a behavior is in agreement with what is present in literature, see for example \cite{Cortesi2017,Cortesi2018}.
In particular, for the \fullthgem{}, an almost constant IBF of about 10\% seems to be reached at sufficiently high gain (above $10^3$).

\begin{figure}[h!]
	\centering
	\includegraphics[width=0.5\textwidth]{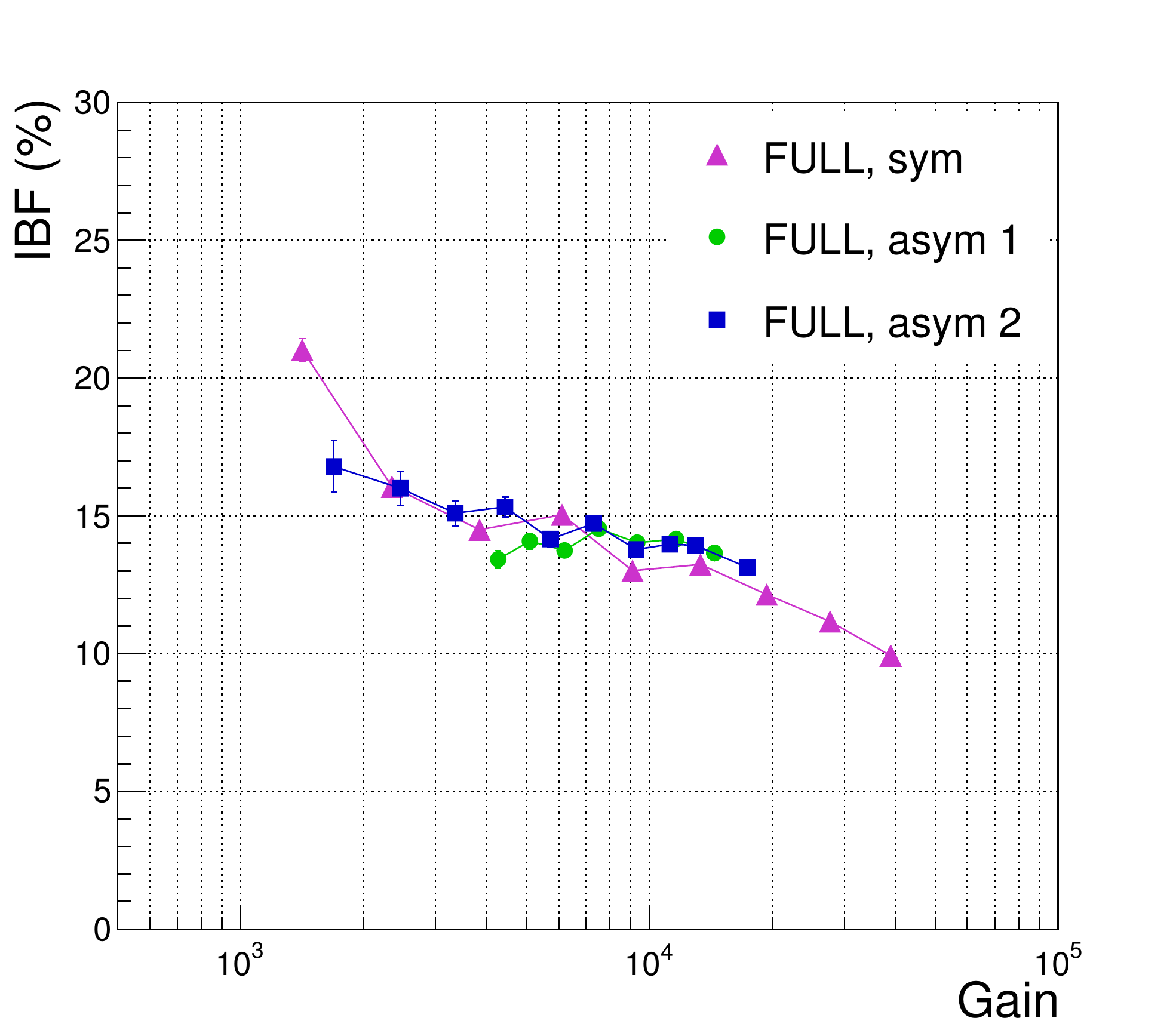}
	\caption{IBF as a function of the gain for one symmetric and two asymmetric configurations for the \fullthgem{} at $P = $10 mbar, \Vind{} = 70 V and \Vdrift{} = 600 V. Sym: V$_{TH1}$ = V$_{TH2}$ = V$_{TH3}$ = 120$-$210 V; Asym 1: V$_{TH1}$ = V$_{TH2}$ = 150$-$190 V, V$_{TH3}$ = 200 V; Asym 2: V$_{TH1}$ = 200 V, V$_{TH2}$ = V$_{TH3}$ = 150$-$195 V.}
	\label{fig:gainasym}
  \end{figure}

In all the above discussed tests the three layers of the M-THGEM foil were set at the same bias (symmetric bias configuration). 
We also studied asymmetric M-THGEM configuration, i.e. configuration where the three layers are set at different voltages, in order to investigate possible effects on the IBF.
The tests were performed with the \fullthgem{} at $P = 10$~mbar, \Vdrift{} $=  600$~V,  \Vind{} $ = 70$~V.
For the sake of clarity, we define V$_{TH1}$, V$_{TH2}$, and V$_{TH3}$ as the voltage difference across the top, middle and bottom layers of the M-\thgem{}, respectively.
In Fig. \ref{fig:gainasym} the symmetric configuration (magenta curve), obtained varying V$_{TH1}$ = V$_{TH2}$ = V$_{TH3}$ between 120 and 210 V, is compared with two asymmetric configurations (blue and green curves).
The first asymmetric configuration (green line) is obtained keeping V$_{TH3}$ to a fixed value of 200~V and increasing V$_{TH1}$ and V$_{TH2}$ from 150 to 190 V (discharge value), maintaining V$_{TH1} = $ V$_{TH2}$.
The second asymmetric configuration (blue line) is obtained keeping fixed V$_{TH1}$ at 200 V and varying V$_{TH2}$ and V$_{TH3}$ from 150 to 195 V (discharge value) under the condition V$_{TH2} = $ V$_{TH3}$.
The three configurations do not show significant differences, therefore we can conclude that the IBF depends on the total gain of the M-THGEM and not on how it is shared among the three THGEM layers.

\subsection{Rate characterization}

\begin{figure}[t]
	\centering
	\includegraphics[width=0.5\textwidth]{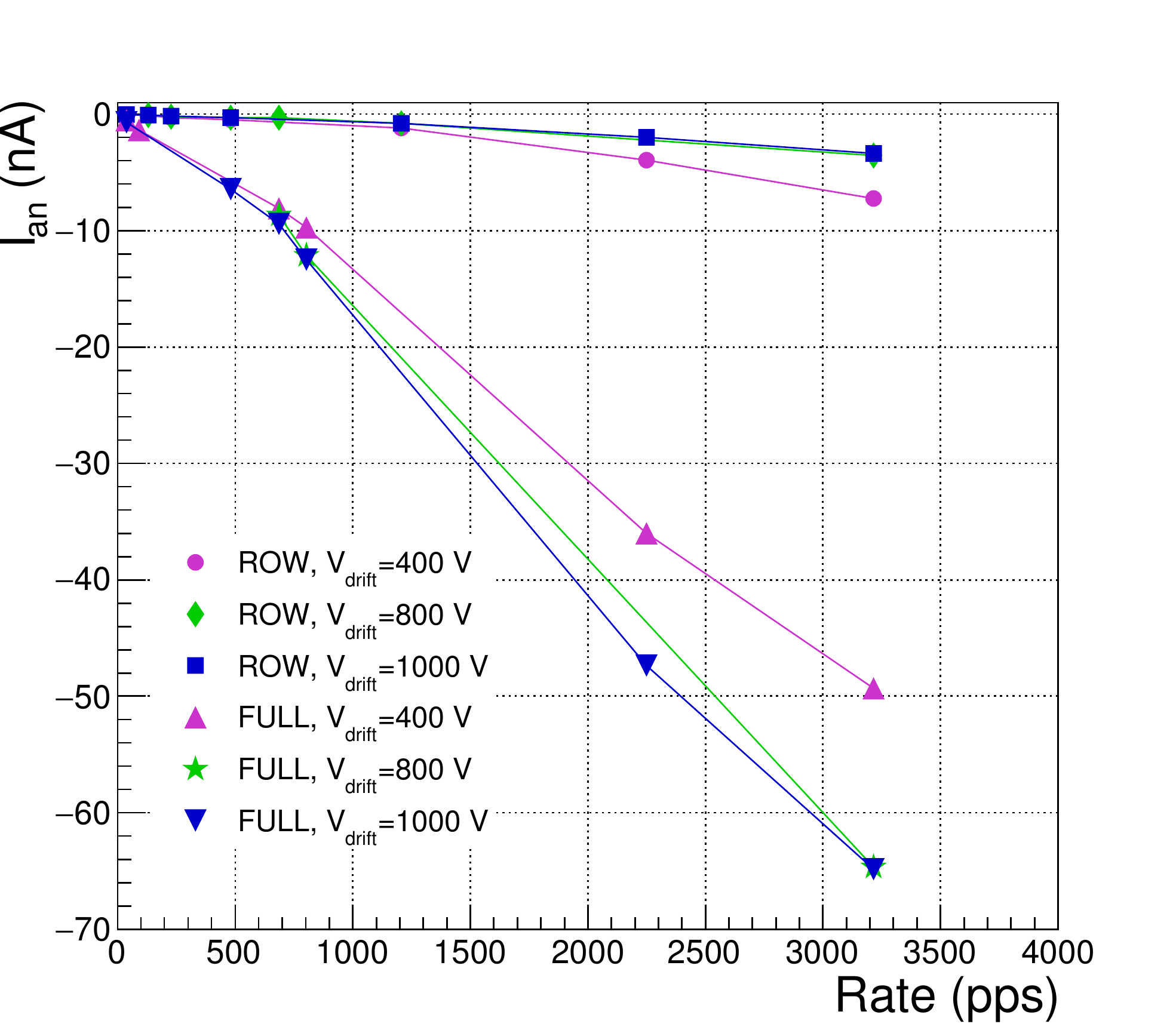}
	\caption{(Color online) Anodic current as a function of the rate of incident particles for the \fullthgem{} and \rowthgem{} at $P = 20$ mbar, \Vind{} = 50 V and \Vthgem{} = 190 V for \rowthgem{} and \Vind{} = 100 V and \Vthgem{} = 205 V for \fullthgem{}, at \Vdrift{}= 400 V (magenta), 800 V (green) and 1000 V (blue).	The data were obtained with $^{18}$O beam and $\alpha$-particle source.
	}
	\label{fig:Rate_ROW_FULL}
  \end{figure}

The effects of the incident particle rate on the M-\thgem{} response were investigated by using the products of scattering of a $^{18}$O beam on Au targets. 
Different rates on the detector ranging from 10 pps to about 3 kpps were obtained using a combination of different beam intensities and target thicknesses.
Since the detector covers a wide horizontal angle ($\sim\ang{15}$) with respect to the scattering center, the rate of incident particles was not uniform along the detector width but was changing of more than two orders of magnitude from one edge to the other, reaching a maximum value of about 300 pps/cm. The rate here reported is the total rate in the detector.
In Fig.~\ref{fig:Rate_ROW_FULL} the plot of the anodic current as a function of the rate at $P=$ 20~mbar is shown. 
For each M-THGEM, three different \Vdrift{} values were 
investigated, keeping the other voltages fixed.
As shown in Sect.~\ref{sec:IBF}, the IBF is strongly dependent on \Vdrift{}. 
Therefore, possible effects in the detector response due to spatial charge should be stronger at low 
\Vdrift{}. The behavior of the anodic current as a function of the particle rate is compatible with a linear one for all the curves.
Therefore, no relevant effects of the rate on the response of the detector are observed up to a rate of about 
3 kpps for all the explored values of \Vdrift{} and for both \fullthgem{} and \rowthgem{} geometries.

\section{Conclusion}\label{sec:conclusion}

In the present work, the characterization of the gas tracker prototype, based on M-THGEM and  developed for the upgrade of the MAGNEX FPD, is reported.
The tests were carried out at the INFN - LNS in Catania in the framework of the NUMEN project.

Two three-layer M-THGEMs with different hole pattern layouts have been used as electron multiplication stage.
In the \fullthgem{}, the holes fill the whole active area, while the \rowthgem{} has five rows of holes parallel to the entrance and exit face of the detector. 

The tests were conducted to study the main features of the prototype, changing the voltages across the tracker regions (drift, THGEM, induction), the gas pressure and the rate of incident heavy-ion particles. They were performed using an accelerated $^{18} \mbox{O}$ beam at 270 MeV and a $^{241}$Am 
$\alpha$-particle source.
The gain and IBF of the prototype, based on the two M-THGEM geometries, have been studied. 
A maximum gain of 4$\times 10^{4}$ was achieved in stable conditions at a pressure of 11 mbar with isobutane gas. 
For the \fullthgem{}-based detector, IBF can be as low as 10\%, whilst for the \rowthgem{} the measured IBF is larger. 
However, as clearly shown in Fig. \ref{fig:Rate_ROW_FULL}, such large IBF values for the \rowthgem{} do not affect the measured detector response at the maximum explored rate of about 3 kpps (300 pps/cm).
Thus, the \rowthgem{} can still be taken in consideration for low-pressure gas tracker for medium to heavy incident ions.
However, the features of the \rowthgem{}-based prototype observed in our study, i.e. the reduced electron transparency, the electric field distortions at the vicinity of the holes and the high measured IBF, lead to the conclusion that for high-rate applications such a geometry is not preferable, compared to more standard geometries such as the \fullthgem{}.

Further systematic tests in which the prototype will be equipped with \fullthgem{} and a pad-segmented readout anode will be undertaken.
Measurements with $\alpha$-particle source are planned in order to study the tracking performance of such a detector in terms of position and angular resolution.
Moreover, in-beam tests will be carried out to evaluate what is the impact of the possible electric field distortions induced by the IBF on the track reconstruction performance.
Tests with other kinds of micro-pattern gas detectors (e.g. Multi-Mesh \thgem{} \cite{Cortesi2018} ) will be taken into consideration in case the IBF values obtained with the present device are too high to allow an efficient track reconstruction.

\section*{Acknowledgments}
This project has received funding from the European Research Council (ERC) under the European Union's Horizon 2020 research and innovation programme (Grant Agreement No. 714625).

The authors acknowledge the project TEBE-FARE (R16HXFTMCT) funded under the call FARE 2016 of Italian Ministry of Education.



\bibliography{mybibfile}


\appendix
\section{} 
\label{sec:appendix}

\setcounter{table}{0}

\begin{table*} [htbp]
	\begin{center}
	\caption{Values of pressure ($P$), \Vthgem{} and \Vdrift{} adopted for the study of \Vind{} for ROW and FULL M-\thgem{} with $\alpha$-particle source.} 
		\renewcommand{\arraystretch}{1.}
		\begin{tabular} {ccccc}
			M-THGEM &  $P$    & \Vind{} range & \Vthgem{} & \Vdrift{} \\
			        & (mbar) &      (V)      &    (V)    &    (V)    \\ 
			\hline
			ROW  &  20	& 	0 - 110 &  220	& 	800   \\
			ROW  &  20	&   0 - 130 &  210	&   800   \\
			ROW  &  30	& 	0 - 110 &  240	& 	800   \\
			ROW  &  42	& 	0 - 170 &  260	& 	700   \\
			FULL &  11	& 	0 - 150 &  170	& 	600   \\
			FULL &  11	& 	0 - 60  &  170	& 	800   \\
			FULL &  11	& 	0 - 60  &  180	& 	800   \\
			FULL &  20	&   0 - 200 &  200	&   1000  \\
			FULL &  30	& 	0 - 220 &  200	& 	1000  \\
			FULL &  30	& 	0 - 220 &  220	& 	1000  \\
			FULL &  30	& 	0 - 220 &  230	& 	1000  \\
			\hline
		\end{tabular}
		\label{tab:ROWTHGEM_vind}
	\end{center}
\end{table*}

\begin{table*} [htbp]
	\begin{center}
	\caption{Values of pressure ($P$), \Vind{} and \Vthgem{} adopted for the study of \Vdrift{} for ROW and FULL M-\thgem{}.} 
		\renewcommand{\arraystretch}{1.}
		\begin{tabular} {ccccccc}
			Ion &   M-THGEM &   $P$    &  \Vind{}  & \Vthgem{} & \Vdrift{} range &   Fig. \\
			&   & (mbar) &    (V)    &    (V)    &    (V)    &  \\ 
			\hline
			$\alpha$  &   ROW  &  10	& 	50  &  180	& 	0 - 800     &    \ref{fig:ion_backflow_FULLandROW_alpha_beam}\\
			$\alpha$  &   ROW  &  20	&   80  &  210	&   100 - 1000   &   \ref{fig:ion_backflow_FULLandROW_alpha_beam}\\
			$^{18}$O  &   ROW  &  20	& 	50  &  190	& 	30 - 1000    &   \ref{fig:ion_backflow_FULLandROW_alpha_beam}\\
			$\alpha$  &   ROW  &  30	& 	70  &  240	& 	0 - 1400    &    \ref{fig:ion_backflow_FULLandROW_alpha_beam}\\
			$\alpha$  &   ROW  &  30	& 	70  &  230	& 	0 - 1400    &   \\
			$\alpha$  &   FULL &  9	& 	50  &  160	& 	200 - 600   &   \\
			$\alpha$  &   FULL &  11	& 	70  &  190	& 	
			0 - 800     &    \ref{fig:ion_backflow_FULLandROW_alpha_beam}\\
			$\alpha$  &   FULL &  11	& 	70  &  170	& 	600 - 850   &    \\
			$\alpha$  &   FULL &  20	&   100 &  205	&   100 - 1200   &   \ref{fig:ion_backflow_FULLandROW_alpha_beam}\\
			$^{18}$O  &   FULL &  20	&   100 &  205	&   400 - 1100   &   \ref{fig:ion_backflow_FULLandROW_alpha_beam}\\
			$\alpha$  &   FULL &  30	& 	120 &  220	& 	0 -
1500    &    \ref{fig:ion_backflow_FULLandROW_alpha_beam}\\
			\hline
		\end{tabular}
		\label{tab:ROWTHGEM_vdrift}
	\end{center}
\end{table*}

\begin{table*} [htbp]
	\begin{center}
	\caption{Values of pressure ($P$), \Vind{} and \Vdrift{} adopted for the study of \Vthgem{} for ROW and FULL M-\thgem{}.} 
		\renewcommand{\arraystretch}{1.}
		\begin{tabular} {ccccccc}
			Ion &   M-THGEM &   $P$    &  \Vind{}  & \Vthgem{} range & \Vdrift{}   &   Fig.\\
			&   & (mbar) &    (V)    &       (V)       &    (V) &   \\ 
			\hline
			$\alpha$  &   ROW  &  10	& 	50   &  140 - 205	& 	200    &    \ref{fig:gain},\ref{fig:IBF_ROW_THGEM}\\
			 $\alpha$  &   ROW  &  21	&   50   &  180 - 225	&   800    &   \ref{fig:gain_beam}\\
			$^{18}$O  &   ROW  &  20	& 	50   &  165 - 205	& 	800    &    \ref{fig:gain_beam}\\
			$\alpha$  &   ROW  &  22	& 	80   &  120 - 220	& 	300    &    \ref{fig:gain},\ref{fig:IBF_ROW_THGEM}\\
			$\alpha$  &   ROW  &  30	& 	70   &  180 - 240	& 	800    &   \\
			$\alpha$  &   ROW  &  32	& 	70   &  170 - 245	& 	400    &   
\ref{fig:gain},\ref{fig:IBF_ROW_THGEM}\\
			$\alpha$  &   ROW  &  42	& 	80   &  220 - 270	& 	700    &   \\
			$\alpha$  &   FULL &  9	& 	50   &  130 - 210	& 	400    &   \\
			$\alpha$  &   FULL &  11	& 	70   &  120 - 210	& 	600    &    \ref{fig:gain},\ref{fig:IBF_ROW_THGEM},\ref{fig:gainasym}\\
			$\alpha$  &   FULL &  11	& 	70   &  V$_{TH3}$ = 200, V$_{TH1,2}$ = 150 - 190	& 	600    &    \ref{fig:gainasym}\\
			$\alpha$  &   FULL &  11	& 	70   &  V$_{TH1}$ = 200, V$_{TH2,3}$ = 150 - 195	& 	600    &    \ref{fig:gainasym}\\
			$\alpha$  &   FULL &  20	& 	100  &  150 - 215	& 	1000   &    \ref{fig:gain},\ref{fig:gain_beam},\ref{fig:IBF_ROW_THGEM}\\
			$^{18}$O  &   FULL &  20	& 	100  &  160 - 210	& 	1000   &    \ref{fig:gain_beam}\\
			$\alpha$  &   FULL &  30	&   120  &  180 - 235	&   1000   &    \ref{fig:gain},\ref{fig:IBF_ROW_THGEM}\\
			\hline
		\end{tabular}
		\label{tab:ROWTHGEM_vthgem}
	\end{center}
\end{table*}

\end{document}